\documentclass[aps,prb,twocolumn,amsmath,amssymb,superscriptaddress,10pt]{revtex4-2}
\usepackage[english]{babel}
\usepackage{graphicx}
\usepackage{bm}
\usepackage{amsmath,amssymb}
\makeatletter
\newsavebox\myboxA
\newsavebox\myboxB
\newlength\mylenA

\newcommand*\xoverline[2][0.75]{%
    \sbox{\myboxA}{$\m@th#2$}%
    \setbox\myboxB\null
    \ht\myboxB=\ht\myboxA%
    \dp\myboxB=\dp\myboxA%
    \wd\myboxB=#1\wd\myboxA
    \sbox\myboxB{$\m@th\overline{\copy\myboxB}$}
    \setlength\mylenA{\the\wd\myboxA}
    \addtolength\mylenA{-\the\wd\myboxB}%
    \ifdim\wd\myboxB<\wd\myboxA%
       \rlap{\hskip 0.5\mylenA\usebox\myboxB}{\usebox\myboxA}%
    \else
        \hskip -0.5\mylenA\rlap{\usebox\myboxA}{\hskip 0.5\mylenA\usebox\myboxB}%
    \fi}
\makeatother
\usepackage[export]{adjustbox}
\usepackage{float}
\usepackage{siunitx}
\usepackage{booktabs}
\usepackage{color}
\usepackage{xcolor}
\usepackage [latin1]{inputenc}
\usepackage{dcolumn}
\usepackage{etoolbox}
\usepackage{url}
\usepackage[colorlinks]{hyperref}
\usepackage[normalem]{ulem}
\usepackage{array}
\newcolumntype{x}[1]{>{\centering\arraybackslash\hspace{0pt}}p{#1}}
\usepackage{multirow}
\usepackage{accents}
\usepackage{braket}

\usepackage{datetime} 
\newcommand{\ms}[1]{\textcolor{teal} {#1}} 
\begin{document}

\title{Antiferromagnetic resonance in $\alpha$-MnTe}


\author{J.~Dzian}
\thanks{These three authors contributed equally}
\affiliation{LNCMI-EMFL, CNRS UPR3228, Univ.Grenoble Alpes, Grenoble, France}
\affiliation{Faculty of Mathematics and Physics, Charles University, Ke Karlovu 5, Prague, CZ-121 16 Czech Republic}
\author{P.~Kuba\v{s}\v{c}\'{\i}k}
\thanks{These three authors contributed equally}
\affiliation{Faculty of Mathematics and Physics, Charles University, Ke Karlovu 5, Prague, CZ-121 16 Czech Republic}
\author{S.~T\'azlar\r{u}}
\thanks{These three authors contributed equally}
\affiliation{Faculty of Mathematics and Physics, Charles University, Ke Karlovu 5, Prague, CZ-121 16 Czech Republic}
\affiliation{Institute of Physics, Academy of Science of the Czech Republic, Cukrovarnick\'a 10, Prague 6, CZ-16253, Czech Republic}
\author{M.~Bia\l{}ek}
\affiliation{Institute of High Pressure Physics, PAS, ul.\ Soko\l{}owska 29/37, PL-01142 Warsaw, Poland}
\author{M.~\v{S}indler}
\affiliation{Institute of Physics, Academy of Science of the Czech Republic, Cukrovarnick\'a 10, Prague 6, CZ-16253, Czech Republic}
\author{F.~Le~Mardel\'e}
\affiliation{LNCMI-EMFL, CNRS UPR3228, Univ.Grenoble Alpes, Grenoble, France}
\author{C.~Kadlec}
\affiliation{Institute of Physics, Academy of Science of the Czech Republic, Na Slovance 2, CZ-182 00 Prague, Czech Republic}
\author{F.~Kadlec}
\affiliation{Institute of Physics, Academy of Science of the Czech Republic, Na Slovance 2, CZ-182 00 Prague, Czech Republic}
\author{M.~Gryglas-Borysiewicz}
\affiliation{Faculty of Physics, University of Warsaw, Pasteura 5, Warsaw, Poland}
\author{K.~P.~Kluczyk}
\affiliation{Faculty of Physics, University of Warsaw, Pasteura 5, Warsaw, Poland}
\author{A.~Mycielski}
\affiliation{Institute of Physics, Polish Academy of Sciences, Aleja Lotnikow 32/46, PL-02668 Warsaw, Poland} 
\author{P.~Skupi\'nski}
\affiliation{Institute of Physics, Polish Academy of Sciences, Aleja Lotnikow 32/46, PL-02668 Warsaw, Poland}
\author{J.~Hejtm\'anek}
\affiliation{Institute of Physics, Academy of Science of the Czech Republic, Cukrovarnick\'a 10, Prague 6, CZ-16253, Czech Republic}
\author{R.~Tesa\v{r}}
\affiliation{Institute of Physics, Academy of Science of the Czech Republic, Cukrovarnick\'a 10, Prague 6, CZ-16253, Czech Republic}
\author{J.~\v{Z}elezn\'y}
\affiliation{Institute of Physics, Academy of Science of the Czech Republic, Cukrovarnick\'a 10, Prague 6, CZ-16253, Czech Republic}
\author{A.-L.~Barra}
\affiliation{LNCMI-EMFL, CNRS UPR3228, Univ.Grenoble Alpes, Grenoble, France}
\author{C.~Faugeras}
\affiliation{LNCMI-EMFL, CNRS UPR3228, Univ.Grenoble Alpes, Grenoble, France}
\author{J.~Voln\'y}
\affiliation{Faculty of Mathematics and Physics, Charles University, Ke Karlovu 5, Prague, CZ-121 16 Czech Republic}
\author{K.~Uhl\'\i\v{r}ov\'a}
\affiliation{Faculty of Mathematics and Physics, Charles University, Ke Karlovu 5, Prague, CZ-121 16 Czech Republic}
\author{L.~N\'advorn{\'\i}k}
\affiliation{Faculty of Mathematics and Physics, Charles University, Ke Karlovu 5, Prague, CZ-121 16 Czech Republic}
\author{M.~Veis}
\affiliation{Faculty of Mathematics and Physics, Charles University, Ke Karlovu 5, Prague, CZ-121 16 Czech Republic}
\author{K.~V\'yborn\'y}
\affiliation{Institute of Physics, Academy of Science of the Czech Republic, Cukrovarnick\'a 10, Prague 6, CZ-16253, Czech Republic}
\author{M.~Orlita}
\email[]{milan.orlita@lncmi.cnrs.fr}
\affiliation{LNCMI-EMFL, CNRS UPR3228, Univ.Grenoble Alpes, Grenoble, France}
\affiliation{Faculty of Mathematics and Physics, Charles University, Ke Karlovu 5, Prague, CZ-121 16 Czech Republic}

\date{\today \ms{{} at \currenttime}}

\begin{abstract}
Antiferromagnetic resonance in a bulk $\alpha$-MnTe crystal is investigated using both frequency-domain and time-domain THz spectroscopy techniques. At low temperatures, an excitation at the photon energy of $(3.5\pm 0.1)$~meV is observed and identified as a magnon mode through its distinctive dependence on temperature and magnetic field. This behavior is reproduced using a simplified model for antiferromagnetic resonance in an easy-plane antiferromagnet. The results of our experiments, when compared to exchange constants established in the literature, allow us to extract the out-of-plane component of the single-ion magnetic anisotropy reaching $(40\pm10)$~$\mu$eV.
\end{abstract}
\maketitle

\section{Introduction}

Manganese telluride (MnTe) is a well-known antiferromagnetic semiconductor~\cite{AllenSSC77} with electric and magnetic properties that make it relevant for various applications, including thermoelectrics~\cite{MuPRM19} or spintronics~\cite{Baltz:2018_a}. A renewed impetus to the research on MnTe -- in particular, on the NiAs-type polymorph, conventionally referred to as $\alpha$-MnTe -- came along with recent experimental observations~\cite{KriegnerNC16,KriegnerPRB17,KluczykPRB24,Hubertpss25, HarikiPRL24,GonzalezBetancourtNJPS24} comprising the anomalous Hall effect (AHE)~\cite{Nagaosa:2010_a}, anisotropic magnetoresistance~\cite{Ritzinger:2023_a}, and the magnetic circular dischroism (MCD) effect at optical frequencies~\cite{Hubertpss25}.  It is also 
the first {\em collinear} magnetic system where splitting
of magnonic bands related to mode chirality~\cite{SmejkalPRL23} has been
experimentally demonstrated~\cite{LiuPRL24}. This splitting is the hallmark of so-called altermagnets~\cite{SmejkalPRX22I,SmejkalPRX22II,BluegelNP25}. The broken parity-time ($PT$) symmetry, characteristic of altermagnets 
(explained in note~\footnote{Combined spatial and time inversion $T$. In addition to this symmetry, translations combined with $T$ also have to be absent among the symmetries. It should be stressed that non-collinear magnets can, of course, also have broken $PT$ symmetry.}),
is manifested by spin splitting of electronic bands~\cite{KrempaskyNature24,OsumiPRB24} that does not rely on spin-orbit coupling (SOC). 

\begin{figure*}
    \includegraphics[width=.65\textwidth,valign=t]{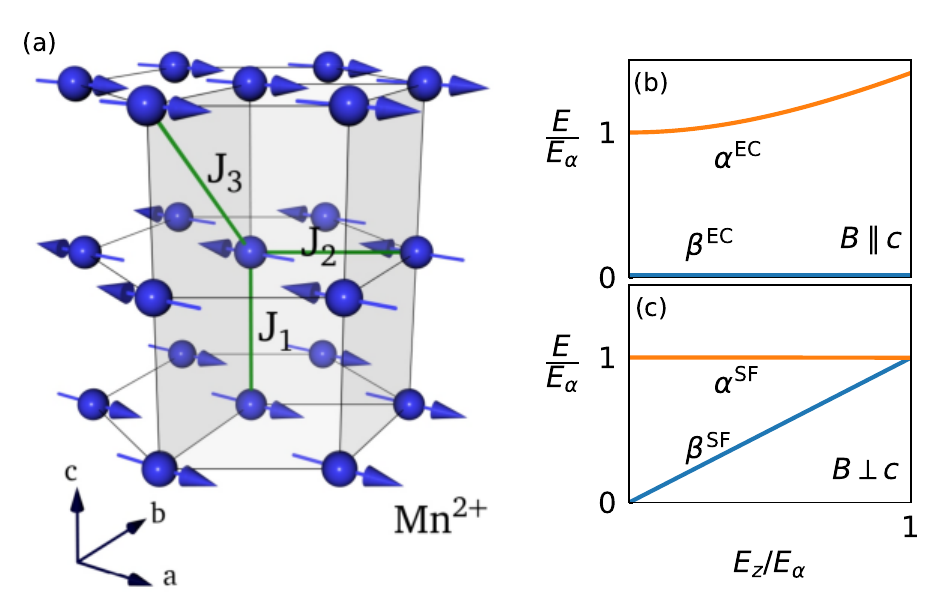} 
    \caption{(a) Magnetic structure of $\alpha$-MnTe with ferromagnetic arrangement of Mn$^{2+}$ spins in each hexagonal plane and antiferromagnetic coupling between adjacent planes. The green lines show the exchange coupling terms in Hamiltonian (\ref{Heisenberg}), the gray volume marks the magnetic unit cell. Panels (b) and (c) show the theoretically expected $B$-dependence of the $\beta$ and $\alpha$ magnon ($k=0$) modes for the magnetic field applied along and perpendicular to the $c$ axis, respectively. For details see Sec.~\ref{Theory}. }
    \label{fig_intro}
\end{figure*}

Bulk $\alpha$-MnTe displays an indirect band gap around 1.3~eV at the liquid helium temperature~\cite{Ferrer-RocaPRB00}, with the conduction band minimum in the center of the Brillouin zone and valence band maxima close to the $A$ point~\cite{JuniorPRB23}. Somewhat higher values, up to 1.5~eV, were reported for thin presumably strained epitaxial layers at room temperature~\cite{KriegnerNC16}. Regarding magnetic structure and properties, $\alpha$-MnTe is an antiferromagnet with an ordering temperature $T_N$ close to 310~K~\cite{Madelung2000}. The magnetic anisotropy is biaxial with a considerably stronger out-of-plane component which aligns the Mn$^{2+}$ spins in the hexagonal planes~\cite{TakemiJPSJ63,KriegnerPRB17,AlaeiPRB25}. Within each plane, the spins are ordered ferromagnetically. The coupling between adjacent planes is antiferromagnetic (Fig.~\ref{fig_intro}a). 

The dispersion of magnon excitations in $\alpha$-MnTe was mapped by means of inelastic neutron scattering experiments and modeled, using linear spin wave (LSW) theory, by Szuszkiewicz et al.~\cite{SzuszkiewiczPRB06}. In their model, they considered an easy-plane type of the antiferromagnetic order and therefore, as is known in NiF$_2$~\cite{MoriyaPR60}, the double degeneracy of magnon dispersion is lifted at the center of the magnetic Brillouin zone ($k=0$ magnons). Experimentally, a single magnon mode at the energy of $(2.6\pm0.8)$~meV, most likely the upper magnon branch, was reported at the $\Gamma$ point in~\cite{SzuszkiewiczPRB06}. In a more recent neutron scattering experiment by Liu et al.~\cite{LiuPRL24}, a bit higher energy of this mode was concluded, close to 4~meV, based on data modeling in the framework of the LSW theory.

The energies and splitting of long-wavelength magnon modes are largely determined by the orientation and strength of single-ion magnetic anisotropy which plays a key role in defining the type and stability of ordered phases, not only in antiferromagnetic materials (see, e.g., \cite{OGradyJAP20, RezendeJAP19}). Valuable insights in magnetic anisotropy can thus be obtained by experimentally probing the $k=0$ magnons. For this purpose, magnetic resonance is the method of choice, as it allows for precise determination of magnon gaps, typically with greater accuracy than inelastic neutron scattering.

In this study, we present antiferromagnetic resonance (AFMR) experiments in the THz spectral range conducted on bulk $\alpha$-MnTe across a wide range of temperatures and applied magnetic fields. The observed behavior is consistent with theoretical predictions for an easy-plane (hard-axis) antiferromagnet, based on the LSW theory. Our modeling allows us to estimate the out-of-plane anisotropy and compare it with the results of previous studies using inelastic neutron scattering~\cite{SzuszkiewiczPRB06,LiuPRL24}.

\section{Theoretical model}
\label{Theory}

The AFMR is a phenomenon widely explored in solid-state physics, approached theoretically at the classical and quantum levels~\cite{KittelPR51,KefferPR52,RezendeJAP19}. It refers to resonant absorption of light in antiferromagnets that emerges due to the precession of exchange-coupled spin sub-lattices subjected to an externally applied magnetic field. Alternatively, the AFMR mode can be viewed as an optically driven excitation of the $k=0$ magnon modes.

Here we summarize theoretical expectations for the AFMR in an easy-plane antiferromagnet, which can be taken as the first approximation for $\alpha$-MnTe. Hence, we neglect the in-plane anisotropy as well as the (higher-order) Dzyaloshinskii-Moriya interaction which are both relevant for $\alpha$-MnTe~\cite{KluczykPRB24}, but they imply energy scales at least one order of magnitude smaller than the out-of-plane anisotropy. 
Similarly, the proposed model, see Appendix~\ref{LSWT} for details, disregards magnon-magnon interactions. Their inclusion should not lead to a qualitative change in the dispersion of magnons, apart from its broadening~\cite{GarciaGaitan25}. Hence, our model should be sufficient for a description of the dispersion profile.

Let us consider the standard Heisenberg-type Hamiltonian, comprising the exchange interactions up to the third nearest neighbors. It includes
the inter-sublattice coupling ($J_{1,3}$) and intra-sublattice coupling ($J_2$), cf. Fig.~\ref{fig_intro}a, the single-ion anisotropy ($D>0$), as well as the Zeeman term: 
\begin{equation}
\begin{split}
\hat{\mathcal H} = J_1 \sum_{{\langle ij\rangle}_1} {\bf S}_i\cdot {\bf S}_j +
J_2 \sum_{{\langle ij\rangle}_2} {\bf S}_i\cdot {\bf S}_j + J_3 \sum_{{\langle ij\rangle}_3} {\bf S}_i\cdot {\bf S}_j \\ + D \sum_i (S_i^z)^2 - g\mu_B \sum_{i}\mathbf B\cdot {\mathbf S}_i\,,
\end{split}
\label{Heisenberg}
\end{equation}
where the $z$ direction is parallel to the $c$ axis and spin $S$ reaches 5/2 for Mn$^{2+}$ ions. The magnetic field has a general orientation, and in our approximation of an easy-plane antiferromagnet, we describe it by the angle $\theta_B$ between $\mathbf B$ and the $c$ axis. 

The above Hamiltonian implies three distinct energy scales: the Zeeman energy $E_Z=g \mu_B B$, the magnetic anisotropy energy $E_{\mathrm{an}}=DS$ (due to both the SOC affecting the band structure and the dipole-dipole interaction~\cite{Correa:2018_a}), and the effective exchange energy $E_J=\sum_{n}J_n {\cal Z}_n S$, where $n$ runs over the neighboring inter-sublattice spins ($n=1$ and 3), with the exchange coupling strength of $J_n$ and the corresponding coordination number ${\cal Z}_n$ (${\cal Z}_1 = 2$, ${\cal Z}_3 = 12$). Let us note that the intra-sublattice coupling $J_2$ does not enter $E_J$ in the macrospin representation, used to calculate the ground state, see Appendix~\ref{LSWT}.

\begin{figure}[t]
\includegraphics[width=1.05\columnwidth,valign=t]{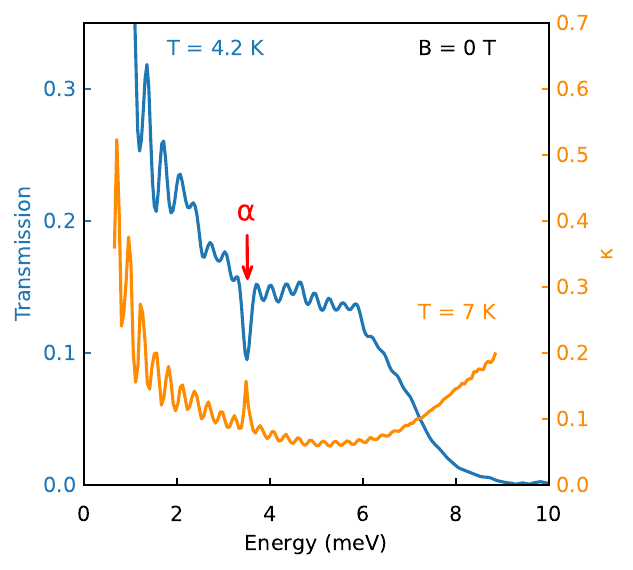} 
\caption{Low-temperature THz transmission spectrum of the bulk $\alpha$-MnTe layer with the thickness of $d=400$~$\mu$m (sample A), measured using the frequency-domain Fourier-transform technique, compared with the extinction coefficient $\kappa$ obtained using the time-domain THz spectroscopy. The vertical arrow indicates the energy of a magnetic excitation identified as the $\alpha$ mode in this work.}
\label{fig_T-lowT}
\end{figure}

The magnon dispersion, including the $k=0$ modes relevant for our optical and magneto-optical experiments, can be obtained using the LSW theory~\cite{SzuszkiewiczPRB06, RezendeJAP19}. In antiferromagnets with easy-plane magnetic anisotropy and two spins per unit cell, such as $\alpha$-MnTe (see Fig.~\ref{fig_intro}a), we expect the degeneracy of the two magnon modes to be lifted throughout the Brillouin zone. At the $\Gamma$ point, the separation between them scales with the strength of the magnetic anisotropy. At non-zero momenta, interestingly, a part of the splitting may be associated with the altermagnetic nature of $\alpha$-MnTe~\cite{SmejkalPRL23,LiuPRL24}. In the following, we will refer to the lower and upper magnon branches at $k=0$ as the $\beta$ and $\alpha$ modes, respectively.    

The LSW theory predicts the energies of both magnon modes, $\hbar\omega_\alpha(B)$ and $\hbar\omega_\beta(B)$, for any strength and direction of the applied magnetic field. Particularly simple analytic expressions are obtained in two limiting cases: for $\mathbf{B} \perp c$ ($\theta_B=\pi/2$), when $B$ exceeds the corresponding spin-flop fields ($B_\mathrm{sf}$ below 4~T, see Refs.~\cite{KriegnerPRB17,BeyCM24}) and for ${\mathbf B}\,\|\,c$ ($\theta_B=0$). In the following, we will call these cases ``spin-flop'' (SF) and ``even canting'' (EC) configurations, respectively. 

In our simplified Hamiltonian for spins in $\alpha$-MnTe~(\ref{Heisenberg}), the $k=0$ magnon energies depend on the applied magnetic field ($E_Z\propto B$) 
as 
\begin{align}
\label{ECa}
\hbar\omega^\mathrm{EC}_\beta & = 0, \\
\hbar\omega^\mathrm{EC}_\alpha & =\sqrt{4E_{\mathrm{an}}E_J+\frac{E_Z^2(1-E_{\mathrm{an}}/E_J)}{(1+E_{\mathrm{an}}/E_J)^2}},  \label{ECb}         
\end{align}
for the EC case (for any magnitude of $B$), while in the SF case, we obtain for $B>B_\mathrm{sf}$ (but $E_Z\le 2E_J$)   
\begin{align}
\label{SFa} 
\hbar\omega^\mathrm{SF}_\beta & =E_Z\sqrt{1-\frac{E_{\mathrm{an}}}{E_J}},\\
\hbar\omega^\mathrm{SF}_\alpha &=\sqrt{4E_{\mathrm{an}}E_J-\frac{E_{\mathrm{an}}E_Z^2}{E_J}}. \label{SFb}  
\end{align}
The case of the SF configuration, but at magnetic fields below $B_\mathrm{sf}$, was treated theoretically, e.g., in Ref.~~\cite{NagamiyaAiP55}. The energies of both magnon modes are expected to depend strongly on the initial angle between magnetic moments and the applied (in-plane) magnetic field. We did not observe such effects in $\alpha$-MnTe, but they were studied in some other easy-plane systems with higher spin-flop fields, e.g., in NiPS$_3$~\cite{JanaPRB23}. Let us also note that, at $B=0$, the above results agree with Eqs.~23 and 24 of Ref.~\cite{RezendeJAP19} in the limit of a vanishing in-plane anisotropy. The nature of $\alpha$ and $\beta$ modes can be visualized in terms of a classical precession of sub-lattice magnetizations (see Fig.~5 in \cite{RezendeJAP19}). In the aforementioned limit, the corresponding trajectories become highly elliptical and linear, respectively.

In Figs.~\ref{fig_intro}b and c, we plot the expected $B$ dependence of the $k=0$ magnon modes. Both energies are normalized by the energy of the $\alpha$ mode at $B=0$ ($E_\alpha$). The realistic ratio $E_{\mathrm{an}}/E_J = 0.003$, justified \emph{a posteriori} in this work, was used. We find distinctively different behaviour in the SF and EC configurations. In the EC case, the $\alpha$ mode undergoes a profound blueshift, approximately quadratic at low magnetic fields, while the $\beta$ mode stays at zero, unless effects due to the considerably weaker in-plane anisotropy are considered, see \emph{e.g.}, Ref.~\cite{RezendeJAP19}. In the SF case, the $\alpha$ mode undergoes a weak redshift with $B$, while the $\beta$ mode approaches the spin resonance of a free electron.  Unfortunately, as discussed in the following, only the $\alpha$ mode is observed in our THz experiments. 

\begin{figure*}[t]
\includegraphics[width=0.95\textwidth,valign=t]{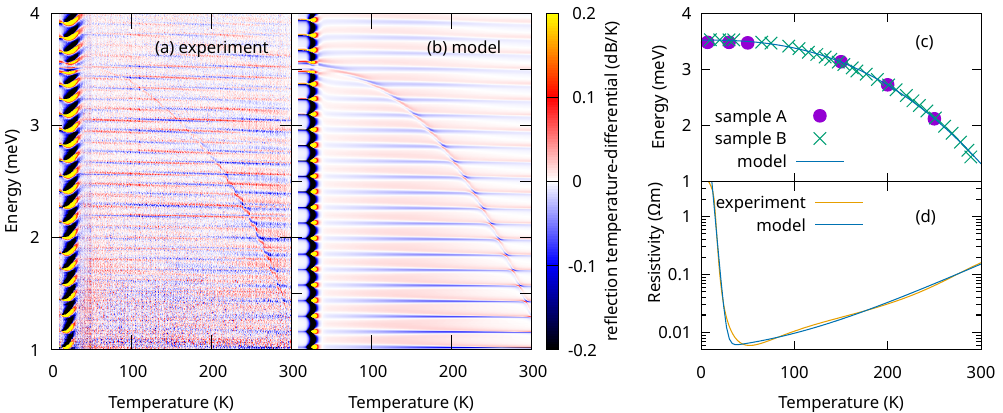} 
\caption{(a) Temperature-differential reflectivity of $\alpha$-MnTe (sample B) as a function of frequency and temperature measured using the time-domain technique, compared with results of the theoretical model (b) explained in the main text and Appendix~\ref{Drude}. (c) Experimentally determined magnon energy as a function of temperature for samples B and A. Observed decrease in energy is compared with 
universality class models implying $(T/T_N)^\zeta$ scaling~\cite{Kobler10}.
(d) Electrical resistivity of sample B as a function of $T$ with the theoretical fit described in Appendix~\ref{Drude}.}
\label{fig_THz-T}
\end{figure*}

\section{Experimental}

Bulk $\alpha$-MnTe samples, studied in this work, were prepared in two batches using different growth methods. Sample A was grown using the self-flux technique. Pure manganese (99.9998\%) and tellurium (99.9999\%) in the molar composition Mn$_{33}$Te$_{67}$ were placed in an alumina (99.95\%) crucible and, together with a catch crucible filled with quartz wool, sealed in a fused-silica tube under vacuum. The sample was heated up to 1050~$^{\circ}$C and then cooled down to 760~$^{\circ}$C for four days. At 760~$^{\circ}$C, the sample was quickly placed in a centrifuge, separating the crystals from the remaining melt. The crystals were flat plates with lateral dimensions of several millimeters and a thickness of several hundred microns. Bulk $\alpha$-MnTe crystals grown under identical conditions had been characterized using powder and single-crystal X-ray techniques, magnetic susceptibility measurements, and XPS spectroscopy in Ref.~\cite{KrempaskyNature24}.

Sample B was produced by the vapor-solid method. Pure tellurium and manganese powders (99. 999\%) were placed at two ends of a quartz ampule, which was sealed under vacuum and heated. After evaporation of tellurium at the temperature of 600~$^{\circ}$C, at about 950~$^{\circ}$C tellurium vapors reacted with manganese, forming irregular shaped crystals with  typical dimensions of several millimeters. The crystals were characterized by X-ray diffraction (both powder and monocrystal one). An equivalent sample had been extensively studied using the electron transport and magnetization methods in Ref.~\cite{KluczykPRB24}.

\begin{figure*}
    \includegraphics[width=.65\textwidth,valign=t]{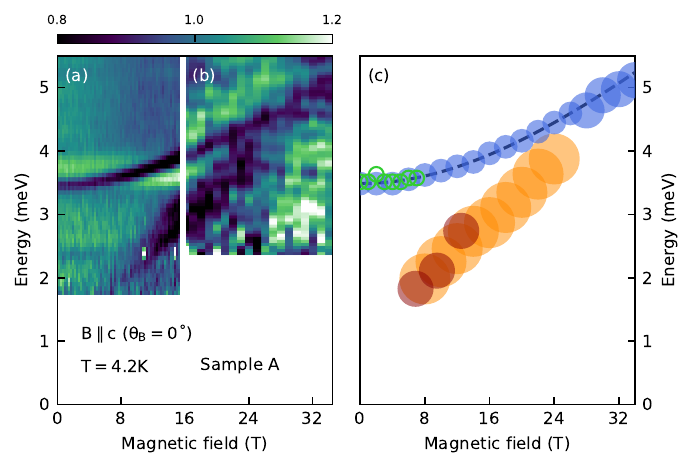} 
    \caption{A color map of magneto-transmission taken on sample A at liquid helium temperature using (a) superconducting and (b) resistive coils, with the magnetic field applied along the $c$ axis. The magneto-transmission $T_B$ is normalized by the transmission averaged over the whole field range. The extracted positions of resonances with a step of 2~T are plotted as blue circles in (c), complemented with data obtained by the time-domain THz and ESR techniques (green and dark orange circles, respectively), see Appendix~\ref{EPRdata} for the latter. The size of the circles corresponds to the error bar. For the dashed line, see the main text.}
    \label{fig_FTIR-B}
\end{figure*}

Four different methods of THz spectroscopy were employed to gather  experimental data presented in this paper: (i) Fourier-transform magneto-spectroscopy at low temperatures (Figs.~\ref{fig_T-lowT}, \ref{fig_FTIR-B} and \ref{fig_FTIR_tilt}); (ii) the time-domain technique employed at $B=0$ (Figs.~\ref{fig_T-lowT} and \ref{fig_THz-T}) in a broad range of temperatures with a complementary set of data collected under magnetic field applied at low temperatures (Fig.~\ref{fig_FTIR-B}c); (iii) magnetic-circular-dichroism (MCD) technique at selected THz frequencies and varying temperatures (Fig.~\ref{fig_MCD}); and (iv) high-frequency electron-spin resonance (ESR) in Figs.~\ref{fig_FTIR-B}c and \ref{fig_ESR}.  
These experimental techniques are described in detail in  Appendix~\ref{Exp}.




\section{Results and discussion}
Let us start the discussion of our experimental results with the transmission spectrum of sample A measured using the Fourier-transform technique at $B=0$ and $T=4.2$~K, see the blue line in Fig.~\ref{fig_T-lowT}. In the THz range, the transmission gradually decreases with photon energy and vanishes above 9~meV due to the onset of strong phonon absorption. In the explored spectral window, the transmission spectrum is weakly modulated by an interference pattern. We observe only a single absorption mode marked by the vertical arrow in Fig.~\ref{fig_T-lowT}. The same excitation is also visible in the extinction coefficient $\kappa$ extracted using the time-domain THz spectroscopy (at $T=7$~K), see the orange line in Fig.~\ref{fig_T-lowT}. As justified \emph{a posteriori}, by the characteristic temperature and magnetic-field dependence, this is the AFMR signal of $\alpha$-MnTe. 
Equivalently, we may call it the $k=0$ magnon mode that corresponds to the upper dispersion branch ($\alpha$ mode). At low temperatures, the mode has an energy of $E_\alpha=\hbar\omega_\alpha^{B=0}=(3.5\pm0.1)$~meV. For comparison, the magnon energy reported for an epitaxial layer of bulk cubic MnTe is larger, reaching nearly 4.3~meV at low temperatures~\cite{SzuszkiewiczAPP97,Szuszkiewiczpssb14}.    

In Fig.~\ref{fig_THz-T}a, we present, in the form of a color plot, temperature-differential reflection spectra of sample B, measured using THz time-domain spectroscopy (see Appendix~\ref{Exp}) in the range of $T=10-300$~K. The reflectivity configuration was used, as the sample opacity in the THz range increases significantly with temperature. This is clearly manifested in the optical data by a profound change in the interference pattern around 30~K (Fig.~\ref{fig_THz-T}a). We interpret this as due to free charge carriers (holes), excited thermally from acceptor states. This is in line with the rapid decrease in dc electrical resistivity observed when the temperature increases from 10 to 50~K (Fig.~\ref{fig_THz-T}d).

Even though the $\alpha$ magnon mode and its characteristic redshift with $T$ is directly traceable in the raw data (Fig.~\ref{fig_THz-T}a), we performed a more detailed analysis to extract its position, width, and strength. To describe the contribution of holes to THz conductivity, thus reducing the number of free parameters in the modeling of our time-domain reflectivity data, we reproduced the temperature dependence of dc resistivity by an empirical function, see Appendix~\ref{Drude} for details.
 
The extracted positions of the magnon resonance are plotted in Fig.~\ref{fig_THz-T}c, complemented by several control points obtained on sample A using an independent time-domain THz setup. With increasing $T$, the observed mode exhibits a pronounced redshift, down to 1.4~meV at room temperature. Such a redshift is characteristic of magnon excitations in magnetic materials when approaching, from below, the ordering temperature, see, e.g., Refs.~\cite{NagaiPR69,ZhangJMMM21}. In Fig.~\ref{fig_THz-T}c, we modeled this behavior using the formula $E_\alpha(T)=E_\alpha[1-c(T/T_N)^\zeta]$, with $\zeta$ and $c$ as the fitting parameters. Consistent with Refs.~\cite{KoblerPBCM05,Kobler10}, this yielded values of $\zeta=(2.5\pm0.1)$ and $c=(0.68\pm0.05)$ for $T_N=310$~K.  Our modeling procedure also provided us with the width and oscillator strength (magnetic susceptibility) of the $\alpha$ mode. At low temperatures, these reach $\gamma_m=(20\pm10)$~$\mu$eV and $\chi=(6\pm2)\times10^{-4}$, respectively. The details of modeling and its results are discussed in Appendix~\ref{Drude}, including the temperature dependence.

Our optical measurements on $\alpha$-MnTe in an external magnetic field are summarized in Fig.~\ref{fig_FTIR-B}. The basic set of the collected data, with $\mathbf B$ applied along the $c$ axis of $\alpha$-MnTe (EC configuration) is presented in panel (a). Magnetic excitation, identified above as the $\alpha$ mode, undergoes a pronounced blueshift that is approximately quadratic in $B$. Despite a lower signal-to-noise ratio, the monotonic blueshift is well visible also in the high-field part of the data, see panel (b), and continues up to the highest magnetic field applied (33~T). The extracted energies of the $\alpha$ mode are plotted in panel (c) together with the low-field points obtained using the time-domain THz spectroscopy and ESR techniques. 

Interestingly, another field-dependent spectral feature emerges at lower photon energies, see Figs.~\ref{fig_FTIR-B}a and b and orange circles in Fig.~\ref{fig_FTIR-B}c. One could invoke the possibility that this extra line is the lower magnon branch ($\beta$ mode). In fact, also the $\beta$ mode may acquire a non-zero energy when the in-plane component of anisotropy is not fully negligible. However, a closer inspection of our data collected on other $\alpha$-MnTe samples from batch A excludes this scenario. In contrast to the $\alpha$ line, which is consistently present in all our measurements, the lower energy line is not manifested in all samples. Presumably, this reflects the varying hole density among the samples studied. In addition, the lower-energy line broadens significantly with $B$. This is considerably different from behaviour of the $\alpha$ line and represents another argument against the magnon origin of this extra line. 

To find an interpretation for this extra line, let us recall the nearly linear dependence of its position on the applied magnetic field, see Fig.~\ref{fig_FTIR-B}c. This may remind us of conventional cyclotron resonance (CR) of free charge carriers~\cite{Ashcroft76}. However, at the same time, the zero-field extrapolation yields a nonzero energy. Such behavior was observed in previous CR studies of weakly bound electrons in bulk semiconductors and their heterostructures, see, e.g.,~Refs.~\cite{KotthausPRL75,WigginsSS90,SeckPBCM95,OshikiriPBCM01}. In fact, the presence of electrons localized at low temperatures is consistent with a sharp drop in electrical resistance observed with increasing $T$, see Fig.~\ref{fig_R-T}. The slope of the extra line would imply a rather high effective mass, $m_{\mathrm{eff}}\approx m_0$. This is, however, greater than the previously reported experimental values ($\approx 0.25m_0$) extracted from zero-field optical studies at liquid nitrogen temperature~\cite{ZanmarchiJPCS67,ZanmarchiJAP68} as well as the ab-initio estimate~\cite{JuniorPRB23}.   

\begin{figure}[t]
\includegraphics[width=.48\textwidth,valign=t]{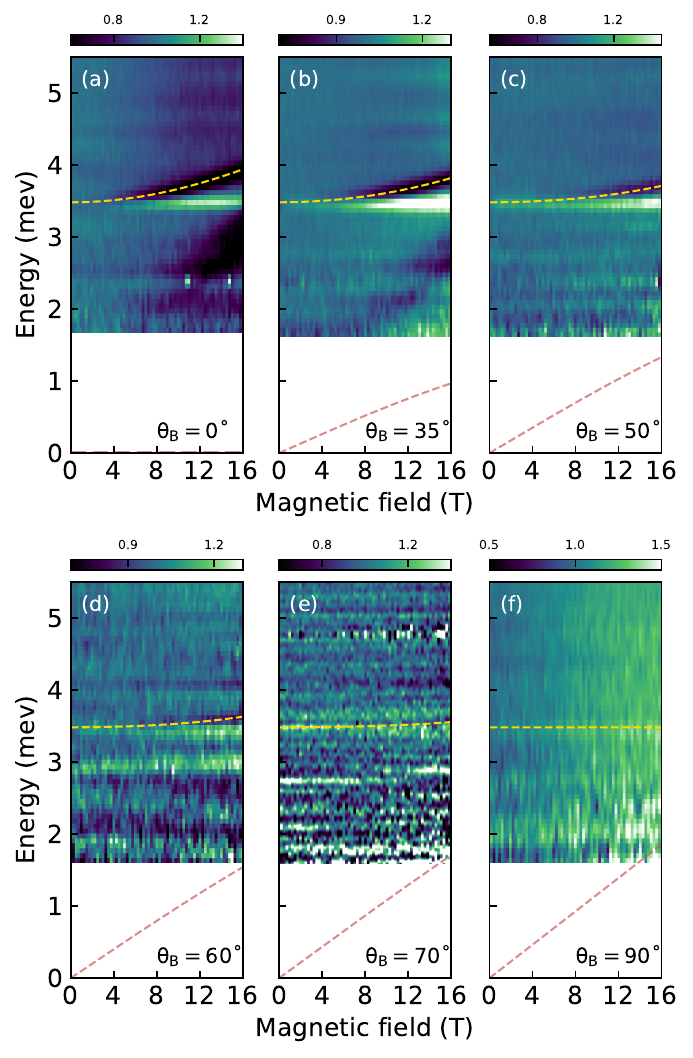} 
\caption{Relative magneto-transmission, $T_B/T_{B=0}$ collected in the Faraday configuration on sample A at liquid helium temperature, plotted as color plots for several values of the angle between the magnetic field and the $c$ axis of $\alpha$-MnTe: $\theta_B=0, 35, 50, 60, 70$ and 90$^{\circ}$ in panels (a) to (f). The data in panel (a) is the same as in Fig.~\ref{fig_FTIR-B}a, but this time normalized by the zero-field transmission. In this way, the dark and light colors correspond to areas of $B$-induced absorption and transmission, respectively. The upper and lower dashed lines are theoretically expected energies of the $\alpha$ and $\beta$ modes, respectively, calculated numerically using the model presented in Sec.~\ref{Theory} for $D=36.4$~$\mu$eV, $g=2.023$ and $E_{an}=91$~$\mu$eV.}
\label{fig_FTIR_tilt}
\end{figure}

Let us now compare the $B$-field dependence of the $\alpha$ mode energy with the proposed theoretical model. The observed blueshift matches well the expectations raised by Eq.~\ref{ECb}. The validity of the theoretical model can be further corroborated using an additional set of data collected at various angles $\theta_B$ between $\mathbf B$ and the $c$ axis. Theoretically, we expect a weakening of the blueshift with increasing $\theta_B$, to gradually approach Eq.~\ref{SFb} with a nearly vanishing $B$-dependence for $E_{\mathrm{an}}\ll E_J$. Experimentally, we have indeed found that the blueshift of the $\alpha$ mode gradually flattens with increasing $\theta_B$. This is illustrated in
Figs.~\ref{fig_FTIR_tilt}a-e where magneto-transmission data collected in the Faraday configuration are presented, this time at varying angle between the applied magnetic field and $c$ axis.  The blueshift of the $\alpha$ mode becomes at the limit of our experimental resolution around $\theta_B=60^{\circ}$. At higher angles, no blueshift is observed, including the limiting case of $\theta_B=90^{\circ} $ which we probed in the Voigt configuration (Fig.~\ref{fig_FTIR_tilt}).  

Having confirmed the validity of our model at the qualitative level, let us proceed with a quantitative analysis. Knowing the energy of the $\alpha$ mode at $B=0$, as well as its blueshift in the EC configuration, Eq.~\ref{ECb} seems to be sufficient to extract both magnetic anisotropy and effective exchange coupling. However, a closer inspection reveals that for a weak magnetic anisotropy, $E_{\mathrm{an}}\ll E_J $, a particular choice of $E_J$ and $E_{\mathrm an}$ parameters leads only to a minor correction of the blueshift which becomes dominantly driven by the Zeeman energy $E_Z$ and the corresponding $g$ factor in it. 

To account for this, we have fitted the experimentally observed blueshift, see the dashed line in Fig.~\ref{fig_FTIR-B}c, using the formula
$(E_\alpha^2 + \xi^2 \mu_B^2 B^2)^{1/2}$, cf.~Eq.~\ref{ECb}. The extracted parameter $\xi=(1.99\pm 0.03)$ then allows us to calculate an approximate effective relation between the magnetic anisotropy and the factor $g$: $D \approx (\sqrt{3}E_\alpha/15) \sqrt{1-\xi^2/g^2}$. It is also worth noting that the parameter $\xi$ corresponds to the lowest possible value of the $g$ factor consistent with our magneto-optical data. The SOC in Mn-based compounds is typically weak and the effective $g$ factor should not significantly deviate from 2. Setting the maximum value at $g=2.1$, we get the following upper limit for the out-of-plane magnetic anisotropy, $D<150$~$\mu$eV. 

Another independent estimate of magnetic anisotropy can be obtained thanks to the AFMR experiment in the SF configuration (${\mathbf B} \perp c$). There, the $\alpha$ mode is expected to redshift approximately by $2(g\mu_B B D)^2/E^3_\alpha$ as one can infer from Eq.~\ref{SFb}. Since no such shift was observed experimentally, we conclude it to be smaller than our experimental resolution (0.5~cm$^{-1}$ @ 16~T). In this way, we obtain the upper bound for the magnetic anisotropy: $D<250$~$\mu$eV. This estimate is less strict, but is consistent with the one based on our AFMR experiments in the EC configuration.

Using exchange parameters available in the literature for $\alpha$-MnTe is another way to obtain an estimate of the magnetic anisotropy. Szuszkiewicz et al.~\cite{SzuszkiewiczPRB06} mapped the magnon dispersion using inelastic neutron scattering and compared their results with the dispersion obtained by the LSW theory, considering a spin Hamiltonian equivalent to Eq.~(\ref{Heisenberg}). They found good agreement with the theory, except for small departures around zone boundaries, and gave solid estimates of the corresponding exchange coupling constants. However, magnetic anisotropy seems to be the least precise parameter deduced, mainly due to the limited data set around $k=0$. 

Taking the exchange parameters $J_1$ and $J_3$ from Szuszkiewicz et~al.~\cite{SzuszkiewiczPRB06} we obtain $E_J \approx 33$~meV  
which is essentially the magnon bandwidth seen in Fig.~5 of that reference. Practically the same value ($E_J\approx 34$~meV) is also found in recent inelastic neutron scattering studies of $\alpha$ -MnTe by Liu et al.~\cite{LiuPRL24}. These effective exchange energies determined experimentally are not far from the value of 27~meV obtained numerically, within the density functional theory (more specifically, DFT+$U$) framework~\cite{MuPRM19}.

Using the experimentally determined $E_J$~\cite{SzuszkiewiczPRB06,LiuPRL24} and the energy of the $\alpha$ mode energy deduced from the optical measurements in this work (see Fig.~\ref{fig_T-lowT}), we conclude that the out-of-plane component of the magnetic anisotropy falls within the interval of $D=E_\alpha^2/(4SE_J)=(40\pm 10)$~$\mu$eV, i.e.,  
$E_{\mathrm{an}}=(100\pm25)$~$\mu$eV. The declared error bar mainly reflects the uncertainty in the exchange parameters taken from the literature.  Having estimated $D$, we may, in turn, evaluate the effective $g$ factor, $g=(2.03\pm0.03)$ based on the AFMR data
collected in the EC configuration, see the discussion above. 

The deduced $D$ may be compared to both experimental and theoretical values determined in the previous work. In inelastic neutron scattering experiments on $\alpha$-MnTe, see Refs. \cite{SzuszkiewiczPRB06} and \cite{LiuPRL24}, the mutually inconsistent values $(22\pm3)$~$\mu$eV and $(48.3\pm 0.5)$~$\mu$eV, respectively, were reported. This suggests that the error bars of microscopic constants deduced in these experiments might be underestimated. A theoretical estimate of the single-ion magnetic anisotropy, $D_{\mathrm{theor}}\approx 20$~$\mu$eV can be found in Ref.~\cite{KriegnerPRB17}. It is also interesting to note that the absolute value of $D$ is comparable to the single-ion magnetic anisotropy in $\alpha$-MnTe lightly doped with lithium. However, the sign seems to be reversed, thus making Li-doped $\alpha$-MnTe easy-axis rather than easy-plane antiferromagnet~\cite{YumnamPRB24,MoseleyPRB22}.  

Knowing the magnetic anisotropy $D$ and the effective $g$ factor experimentally, we may test the observed magneto-optical response for various orientations
of the magnetic field, see color plots in Fig.~\ref{fig_FTIR_tilt}, against theoretical expectations. Although the theoretical model in Sect.~\ref{Theory} only provides us with simple analytical formulae (Eqs.~\ref{ECa}-\ref{SFb}) for special cases, $\theta_B=0$ and $\pi/2$, the numerical solution can be found for any $\theta_B$. The theoretically expected energies of both $\alpha$ and $\beta$ modes have been plotted as dashed lines in Fig.~\ref{fig_FTIR_tilt}. The theoretical model compares well with our experimental data. Indeed, the blueshift of the $\alpha$-mode becomes below our experimental resolution for angles $\theta_B > 60$~deg. 

\begin{figure}[ht]
\includegraphics[width=.4\textwidth,valign=t]{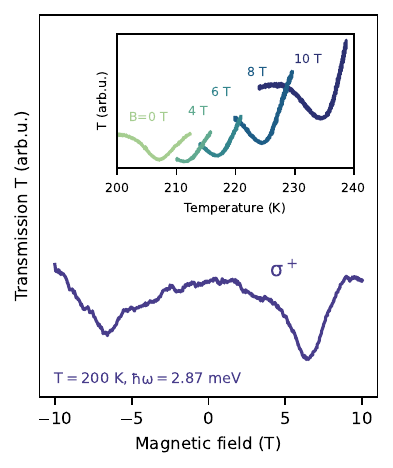} 
\caption{Magneto-transmission of sample A, probed by the THz/FIR laser. The magnetic-field sweep ($\mathbf B \| c$) at a temperature of $T = 200$~K was obtained using circularly polarized radiation with a photon energy of $\hbar\omega = 2.87$~meV. The temperature sweeps, shown in the inset, were measured at several magnetic fields using a linearly polarized laser beam with a photon energy of 2.71~meV.}
\label{fig_MCD}
\end{figure}

Let us close the discussion with another experimentally probed characteristic of the observed $\alpha$ mode, 
its partial circular polarization at $B>0$, see Fig.~\ref{fig_MCD}. It was observed using the THz MCD technique, which operates at a fixed laser frequency with a well-defined circular polarization, and follows the intensity of the transmitted radiation as a function of $B$. The partial circular polarization of the $\alpha$ mode is manifested by a profound difference in the observed absorption when the light propagates parallel or antiparallel to the direction of the magnetic field. 

Magnetic circular dichroism in $\alpha$-MnTe is a complex phenomenon, sensitive to particular symmetries present (or broken), see, e.g.~Ref.~\cite{Hubertpss25}. Nevertheless, the partial circular polarization of the $\alpha$ mode at $B>0$ 
could either be
 an effect of the $B$-induced magnetization or a consequence of the dynamics of magnetic moments (which move in elliptical rather than circular trajectories typical for easy-axis systems). In the ${\mathbf B}\| c$ configuration, the former appears due to spin canting away from the hexagonal plane (the EC regime). The angle between the spin and hexagonal planes reaches a few degrees in the highest applied $B$, $\arcsin[E_Z/(2E_J)]$ {\color{orange} (\ref{GSangles})}, for the expected strength of the exchange coupling~\cite{SzuszkiewiczPRB06}. Note that the experiment with circularly polarized radiation was performed at higher temperatures ($T=200$~K) at which it was possible to achieve resonance between the field-dependent $\alpha$ mode and one of the (discrete) emission lines of the gas laser. This temperature-driven tuning of the magnon energy, at several values of the magnetic field, is demonstrated in the inset of Fig.~\ref{fig_MCD}. 

\section{Conclusions}

A long-wavelength ($k=0$) magnon mode was identified in the THz response of bulk $\alpha$-MnTe using the experimentally observed dependence of its energy on the applied magnetic field and temperature. The observed behavior aligns well with predictions from a simplified model of an easy-plane (hard-axis) antiferromagnet, developed within the framework of linear spin-wave theory. Its energy at low temperatures, 3.5~meV, ranks $\alpha$-MnTe among antiferromagnetic materials with moderate magnon gap (smaller than in some uniaxial systems~\cite{McCreary:2020_a,Csontosova:2023_a} 
but still appreciable). On the other hand, the magnon remains observable even at room temperature, which is close to $T_N$ of hexagonal MnTe.
Combining our experimental results with the values of the exchange parameters published earlier, we determined the out-of-plane component of the magnetic anisotropy as $D=(40\pm 10)$~$\mu$eV and the effective $g$ factor as $g=(2.03\pm0.03)$. These results are characteristic of bulk hexagonal MnTe and may serve as reference values for
this material grown in the form of a thin film, where properties, such as the single-ion anisotropy, and consequently also magnetization, may be significantly impacted by the strain~\cite{BeyCM25,GonzalezBetancourtNJPS24}.

\begin{acknowledgments}
We acknowledge discussions with M. Potemski and D.~Pfannkuche. The work was supported by the Czech Science Foundation, project No. 23-04746S
and TERAFIT project No. CZ.02.01.01/00/22$\_$008/0004594 funded by OP JAK, call Excellent Research. Additional magnetometry on our bulk samples was performed by O. Kaman. J.D. and M.O. acknowledge the support received through the ANR-22-EXES-0001 (project CEQAS). Authors acknowledge the support of the LNCMI-CNRS, a member of the European Magnetic Field Laboratory (EMFL). Supplemental experiments were carried out in MGML (mgml.eu), supported by the Ministry of Education, Youth and Sports, the Czech Republic within the program of the Czech Research Infrastructures (project No.~LM2023065).
 P.K. and L.N. acknowledge funding from the Grant Agency of the Charles University (grants No. 166123 and SVV$-$2024$-$260720). M.B. acknowledges the funding from the SONATA BIS-13 2023/50/E/ST3/00584 grant of the National Science Centre of Poland.
\end{acknowledgments}

\appendix

\section{Linear spin wave theory}
\label{LSWT}

The classical ground state configuration has to be determined first.
Writing the classical energy ${\cal E}$ associated with the Hamiltonian (\ref{Heisenberg}) in the macrospin representation we get for $\mathcal{E}/S$:
$$
E_J \mathbf{m}_1 \cdot \mathbf{m}_2+E_{\mathrm{an}} \left[(\mathbf{z}\cdot\mathbf{m}_1)^2 + (\mathbf{z}\cdot\mathbf{m}_2)^2 \right]-E_Z \mathbf{b}\cdot (\mathbf{m}_1+\mathbf{m}_2),
$$
where $\mathbf{m}_i$ denotes unit vectors corresponding to the respective sublattice magnetizations. The external magnetic field is $\mathbf{B}=\mathbf{b}B$, and without loss of generality, the unit vector $\mathbf{b}$ is taken to be in the plane $\{\mathbf{x},\mathbf{z}\}$. The quantities $E_{\mathrm{an}}$, $E_Z$ and $E_J$ are defined in the main text. 

Minimizing ${\cal E}$, we obtain for $E_Z \leq 2 E_J$ a ``canted spin-flop'' configuration with polar angle $\theta$, shared by both sublattices, and azimuthal angle $\phi$ taken from $\mathbf{x}$-axis as: $\phi_1=-\phi_2=\phi:$   
\begin{align}
\label{GSangles}
\cos\theta&=\frac{E_Z\cos\theta_B}{2\left(E_J+E_{\mathrm{an}}\right)}\approx \frac{E_Z\cos\theta_B}{2E_J} \\ \nonumber
\cos\phi&=\frac{E_Z\sin\theta_B}{2E_J \sin\theta}.
\end{align}
%
We express spin operators in the quantum spin Hamiltonian (\ref{Heisenberg}) in their respective sublattice basis $S^{sl}$, such that $S^{sl,z}_{li} = \mathbf{S}_{li}\cdot \mathbf{z}^{sl}_l$, with $\mathbf{z}^{sl}_l=\mathbf{m}_l$ and these operators are then mapped to the bosonic annihilation and creation operators via Holstein-Primakoff transformation (in its semiclassical limit):
\begin{align*}
&S^{sl,+}_{1i} \stackrel{S\gg 1}{\approx} (2S)^{1/2}a_i,\quad S^{sl,-}_{1i}\stackrel{S\gg 1}{\approx}(2S)^{1/2}a^\dagger_i, \\
&S^z_{1i} = S-a_i^\dagger a_i,
\end{align*}
and similarly for the second sublattice $S^{sl}_{2i}\rightarrow b_i^\dagger,b_i$. Only terms quadratic in bosonic operators are taken into account within the LSW theory, i.e., all terms with three or more operators are neglected (single operator terms do not have any effect on the resulting magnon energies). 
The Fourier-transformed magnon operators $a_k, a_k^\dagger$ allow us to write a quadratic Hamiltonian
${\cal H}_{ab}$ comprising four terms.
\begin{align*}
\sum_{\mathbf{k}}
A_{\mathbf{k}}& (a^\dagger_{\mathbf{k}}a_{\mathbf{k}}+b^\dagger_{\mathbf{k}}b_{\mathbf{k}}) + B_{\mathbf{k}} (a_{\mathbf{k}}b_{-\mathbf{k}}+a^\dagger_{\mathbf{k}}b^\dagger_{-\mathbf{k}}) + \\
+&
\left[
\frac{1}{2} C_{\mathbf{k}}(a_{\mathbf{k}}a_{-\mathbf{k}}+b_{\mathbf{k}}b_{-\mathbf{k}})+h.c. \right]
+ \left[D_{\mathbf{k}}a^\dagger_{\mathbf{k}}b_{\mathbf{k}}+h.c.\right].
\end{align*}
Here we introduce the following  
abbreviations 
\begin{figure}[H]
\begin{align*}
A_{\mathbf{k}}&= E_Z  (\cos\theta\cos\theta_B+\sin\theta\cos\phi\sin\theta_B) \\
&-E_{\mathrm{an}}\frac{1}{2}(1+3\cos2\theta)
-\sum_{n_i}E_{j}^{n_i} (1-\gamma_{\mathbf{k}}^{n_i}) \\
&-\sum_{n_c}E_{j}^{n_c}(\cos^2\theta+\cos2\phi\sin^2\theta)\\
B_{\mathbf{k}} &= \sum_{n_c} E_j^{n_c} \sin^2\theta\sin^2\phi\gamma_{\mathbf
{k}}^{n_c}\\
C_{\mathbf{k}} &= E_{\mathrm{an}} e^{2i\phi}\sin^2\theta \\
D_{\mathbf{k}} &= \sum_{n_c} E_j^{n_c} 
e^{2i\phi}(\cos\phi-i\cos\theta\sin\phi)^2\gamma_{\mathbf{k}}^{n_c}
\end{align*}
\end{figure}
where
$$
E_j^n=J_n{\cal{Z}}_nS,\ \gamma_{\mathbf{k}}^n=
{\cal{Z}}_{n}^{-1}
\sum_{j_n} 
e^{i\mathbf{k}\cdot\mathbf{r}_{j_n}},
$$
with $n\in\{n_c,n_i\}$ and $n_i$ indexing intra-sublattice (within one hexagonal plane) and $n_c$ inter-sublattice exchange parameters $J_n$, ${\cal{Z}}_n$. Note that
$E_J=\sum_{n_c} E_j^{n_c}.$ 

Using Bogoliubov transformation to diagonalize $\mathcal{H}_{ab}$ 
$$
\mathcal{H}_{ab} \stackrel{BT}{\longrightarrow} 
\mathcal{H}_{\alpha\beta} =
\sum_{\mathbf{k}} \omega_{\alpha\mathbf{k}}\alpha^\dagger_{\mathbf{k}}\alpha_{\mathbf{k}} + \omega_{\beta\mathbf{k}}\beta^\dagger_{\mathbf{k}}\beta_{\mathbf{k}}
$$
we arrive, in the end, at
\begin{align*}
\omega_{\beta/\alpha,\mathbf{k}}^2 &= A_{\mathbf{k}}^2-B_{\mathbf{k}}^2-|C_{\mathbf{k}}|^2+|D_{\mathbf{k}}|^2 \mp \sqrt{4|E|_{\mathbf{k}}^2+F_{\mathbf{k}}^2}, \\
E_{\mathbf{k}}&=B_{\mathbf{k}}C_{\mathbf{k}} - A_{\mathbf{k}}D_{\mathbf{k}},\quad F_{\mathbf{k}}= C_{\mathbf{k}}D_{\mathbf{k}}^* - C_{\mathbf{k}}^*D_{\mathbf{k}}. 
\end{align*}
Setting $k=0$ and appropriate $\mathbf{B}$ (out-of-plane or in-plane), we recover Eqs.~(\ref{ECa},\ref{ECb}) for $\theta_B=0$ and Eqs.~(\ref{SFa},\ref{SFb}) for $\theta_B=\pi/2$. This derivation largely follows Ref.~\cite{RezendeJAP19}.


\section{Temperature dependence of electrical resistivity}
\label{transport}
\begin{figure}[t]
\includegraphics[width=.45\textwidth,valign=t]{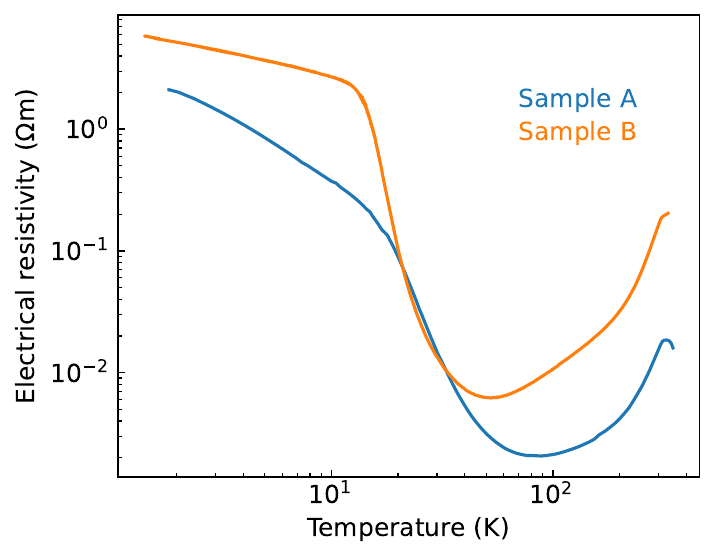} 
\caption{Resistivity dependence on temperature for samples A and B with carrier freeze-out evidenced at low temperatures. The experiments were carried out using the conventional lock-in technique at low frequencies. The profile of $R(T)$ curve qualitatively match overall transmission of $\alpha$-MnTe samples in the THz range, as discussed in the main text.}
\label{fig_R-T}
\end{figure}

The temperature dependence of the electrical resistivity for samples A and B is plotted in Fig.~\ref{fig_R-T}. Qualitatively, both samples show the same behaviour with increasing $T$. A sharp decrease in resistance, with the minima below 100~K, followed by a gradual increase at higher temperatures. At high temperatures, the resistance of sample B appears to be approximately one order of magnitude higher than that of sample A. Assuming that the scattering of free carriers at high temperatures is dominated by intrinsic mechanisms (optical phonons), the difference in the resistance values may suggest a higher carrier density in sample A.  

In terms of experimental details, the resistivity of sample A was measured using a PPMS system, with a maximum current of 100~$\mu$A, using low-frequency ac technique (4.5~Hz), using four contact made by silver epoxy. The resistivity of sample B was measured in a He cooled cryostat with a variable temperature inset. Alternating current of 10~$\mathrm{\mu}$A and frequency of 14.1 Hz was applied to a millimeter-size bar and longitudinal voltage drops were detected with lock-in voltmeters. 

\section{Description of experimental techniques}
\label{Exp}

\emph{Fourier transform magneto-spectroscopy:} Nonpolarized radiation from a mercury lamp was analyzed by the Bruker Vertex 80v spectrometer and delivered, via light-pipe optics, to the sample which was kept at $T=4.2$~K in the helium exchange gas and placed inside a cryostat in a superconducting or resistive coil, below and above 16~T, respectively. In the superconducting coil, the radiation was then detected by a composite bolometer placed just below the sample. In this case, both Faraday and Voigt configurations were possible, with the light propagating along or perpendicular to the applied magnetic field, respectively. In the resistive coil, measurements were made only in the Faraday configuration. In order to reduce the noise level induced by the cooling water, the sample was placed on a mirror and the radiation was detected by an external bolometer. The signal in this latter configuration corresponds to a double-pass transmission. For the absolute transmission measurement in Fig.~\ref{fig_T-lowT}, the sample was mounted on a rotation stage that enabled reference using an open aperture.

\emph{Time-domain THz spectroscopy:} Experiments using this method were carried out in reflection and transmission geometries. Reflection measurements of sample B were performed at an incidence angle of about 9$^{\circ}$, using a THz optical setup with four off-axis parabolic mirrors. We used a photoconductive source and detector antennae, controlled with a standard THz time-domain spectrometer (Toptica). The sample was placed in an optical cryostat with quartz windows, and reflection spectra were collected as a function of the sample temperature in the range of 10-300 K. We calculated temperature-differential reflection spectra that show the $\alpha$ magnon mode because its frequency softens with increasing temperature. These
spectra reveal strong interactions of the magnon with the modes of a Fabry-Perot cavity, formed due to the parallel-plane shape of the sample~\cite{BialekAPL22}. Due to a relatively large sample B thickness (989 $\mu$m), the cavity modes are dense in the frequency domain. 

The time-domain THz experiments in the transmission geometry were performed in two custom-made THz setups. For measurements without magnetic field, broadband pulses (0.2-3~THz) were generated by a photoconducting antenna and detected by electro-optic sampling in a ZnTe crystal. The sample was cooled in an optical cryostat (Optistat, Oxford Instruments) from 300 down to 7~K. For more details on the setup, see Ref.~\cite{BlumenscheinJAP20}. For measurements in a magnetic field, broadband terahertz pulses were generated and detected using commercial fiber-coupled photoconductive switches operating with a femtosecond optical fiber laser system (TeraSmart, Menlo Systems GmbH). The sample was placed in an Oxford Instruments Spectromag He bath cryostat fitted with mylar windows and a superconducting coil, allowing for the application of a magnetic field of up to 7~T, and cooling the sample down to 2 K. The measurements were performed in the Faraday geometry.


\emph{THz magnetic-circular dichroism (THz-MCD):} Temperature and magnetic-field-dependent transmittance measurements with a circularly polarized THz beam were performed in the Faraday geometry. The optically pumped THz/FIR gas laser (Edinburgh Instruments FIRL100) produced a highly monochromatic, linearly polarized THz beam, which was converted into a left-handed or right-handed circularly polarized state using a broadband tunable terahertz retarder~\cite{TesarRSI18}. The cyclotron resonance absorption of a high-quality two-dimensional electron gas in the GaAs quantum well is measured to ensure pure circular polarization. The magneto-optical cryogenic system (Oxford Instruments Spectromag SM4000) provides temperatures from 2 to 300~K and magnetic fields up to 10~T in both polarities. The relative transmittance is given as a ratio of the signal transmitted through the sample and detected by a liquid-helium-cooled bolometer to the signal monitoring the laser output. An additional set of data (see the inset of Fig.~\ref{fig_MCD}) was collected using the described setup, but with linearly polarized radiation instead, to illustrate the fine tuning of the $\alpha$ magnon energy with $T$.

\emph{High-field high-frequency electron spin resonance (ESR):} The explored sample was placed in an open cavity, in the Faraday configuration, in a superconducting coil and kept at low temperatures ($T=5$~K) in the variable temperature insert. The quasi-monochromatic radiation ($\hbar\omega= 1.8$, 2.1 and 2.7 meV were used) was generated via an amplification-multiplication chain of a basic 9.2 GHz synthesized frequency and delivered to the sample by means of quasi-optics. The magnetic-field-modulation technique was employed. The double-transmitted radiation was detected by an external bolometer. For more details, see Ref.~\cite{BarraAMR06}.

\section{Modeling of time-domain THz data}
\label{Drude}

The reflection spectra obtained on sample B by the time-domain technique were modeled using the transfer matrix formalism~\cite{Mackay20}. 
Following the experimental configuration, we considered a plane-parallel slab of $\alpha$-MnTe (the thickness of 989~$\mu$m) placed on a highly reflective metallic layer. The dielectric part of the optical response was described as $\epsilon=\epsilon_{bg}[1-\omega_p^2/(\omega^2+i\omega\gamma_p)]$, where $\omega_p=[Ne^2/(\epsilon_0\epsilon_{\mathrm{bg}}m_{\mathrm{eff}})]^{1/2}$ is the screened plasma frequency. It describes the response due to free carriers (holes) with the density $N$, effective mass $m_{\mathrm{eff}}$, and scattering rate $\gamma_p$. To describe the magnon resonance, magnetic susceptibility was expressed using a Lorentz-type model, $\mu=1+\chi\omega_m^2/(\omega_m^2-\omega^2-i\omega\gamma_m)$, where $\chi$ stands for the oscillator strength of the AFMR, $\omega_m$ is the magnon energy and $\gamma_m$ is the width of the AFMR line.

To reduce the number of free parameters in this optical model, we started with the analysis of the temperature dependence of the dc electrical resistivity (Fig.~\ref{fig_THz-T}d). First, we assumed that this resistivity can be described using a standard Drude-type formula $\rho=m_{\mathrm{eff}}\gamma_p/(Ne^2)$. Second, we used the empirically constructed function: $\rho(T)=a(1+b e^{T/T_b})(1+de^{-T/T_d})$ to reproduce the experimentally observed temperature dependence of electrical resistivity. Good agreement was found for the following parameters: $a=0.39(5)$~$\Omega$m, $b=0.28(5)$, $T_b=62(5)$~K, $d=9.1(5)\times10^4$ and $T_d=2.4(5)$~K, see the solid line in Fig.~\ref{fig_THz-T}d. In the range of 10-50~K, this function describes well the rapid drop in resistivity due to the fast increase in the hole density, $N(T)\propto 1/(1+de^{-T/T_d})$. On the other hand, the formula reproduces the gradual increase in resistivity above 50~K, driven by the decreasing mobility of the holes, $\gamma_p(T)\propto (1+b e^{T/T_b})$. To separate the values of $N(T)$ and $\gamma_p(T)$ in the Drude formula, the carrier density at $T=50$~K was anchored at $5\times 10^{16}$ cm$^{-3}$  -- the value typically observed in Hall experiments in explored samples of $\alpha$-MnTe. Here we assumed that the effective mass is close to that of the bare electron ($m_{\mathrm{eff}} = m_0$). At $T=50$~K, this results in $\hbar\omega_p=(2\pm1)$ meV and $\gamma_p=(30\pm15)$~meV. The latter value shows that absorption due to free charge carriers is flat in the THz range, thus creating a smooth background response for a relatively sharp magnon resonance.

\begin{figure}[t]
\includegraphics[width=.49\textwidth,valign=t]{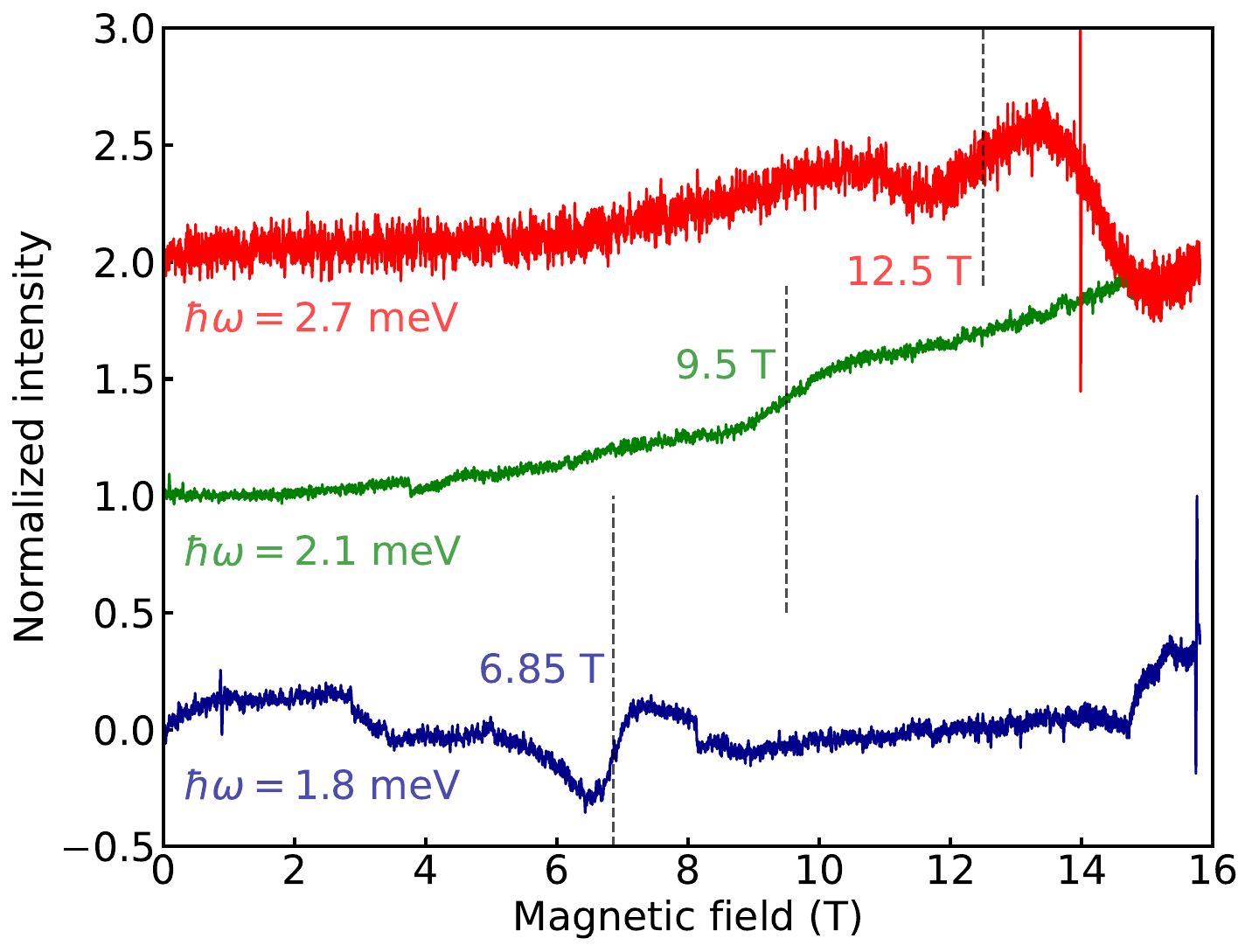} 
\caption{Dataset of $B$-derivative of magneto-transmission measured by the ESR technique. The extracted resonance positions, see vertical dashed lines with labels, are shown as dark orange points in Fig.\ref{fig_FTIR-B}c.}
\label{fig_ESR}
\end{figure}

In addition to the temperature dependence of the magnon energy, presented and discussed in the main text (Fig.~\ref{fig_THz-T}c), our modeling procedure also provided us with the width and strength of the AFMR, including their temperature dependence. The width appeared to be well reproduced by the effective formula: $\gamma_m=\gamma_m^0(1+T[K]/100)$, where $\gamma_m^0=(20\pm10)$~$\mu$eV. Regarding the oscillator strength, its temperature dependence can be traced directly in the experimental data (Fig.~\ref{fig_THz-T}a) through the splitting of the magnon-polariton modes. These are formed when the AFMR crosses subsequent modes of the cavity formed by the sample slab itself. These avoided crossings are related to the vacuum Rabi splitting, which is directly proportional to $\sqrt{\chi}$~\cite{Bialek21, Boventer23}. In our modeling, only the low-temperature part of the data can be reproduced with a constant oscillator strength. At $T> 35$~K, the oscillator strength drops significantly -- by an order of magnitude -- and partly recovers only above about 100~K. Hence, it seems that the magnon oscillator strength and the dc resistivity have a related temperature dependence. One may even invoke the possibility that the magnon oscillator strength is related to the density of thermally excited holes. However, further discussion goes beyond the scope of this paper. For our modeling, we used a heuristic function $\chi(T)$: a constant low-temperature value of $6\times10^{-4}$ that continuously changes into a $\chi(T)\propto R(T)/T^3$ profile in the range of $T=35-300$~ K. The Fabry-Perot pattern also allowed us to establish the background dielectric function that varies smoothly with $T$, $\epsilon_{\mathrm{bg}}(T) = 20.6+i0.1+8\cdot10^{-4}(1+i)T[\mathrm{K}]$.

\section{High-field high-frequency electron spin resonance - dataset}
\label{EPRdata}

A complementary experiment was carried out on sample A using the ESR technique. The experimental data collected at $T=5$~K are plotted in Fig.~\ref{fig_ESR}. There, we plot the $\alpha$-MnTe microwave transmission as a function of the applied magnetic field at three selected microwave frequencies ($\hbar\omega=1.8, 2.1$ and 2.7~meV), in the form of $B$-derivatives. Vertical dashed lines mark the positions of the relevant resonances.

%


\begin{thebibliography}{60}%
\makeatletter
\providecommand \@ifxundefined [1]{%
 \@ifx{#1\undefined}
}%
\providecommand \@ifnum [1]{%
 \ifnum #1\expandafter \@firstoftwo
 \else \expandafter \@secondoftwo
 \fi
}%
\providecommand \@ifx [1]{%
 \ifx #1\expandafter \@firstoftwo
 \else \expandafter \@secondoftwo
 \fi
}%
\providecommand \natexlab [1]{#1}%
\providecommand \enquote  [1]{``#1''}%
\providecommand \bibnamefont  [1]{#1}%
\providecommand \bibfnamefont [1]{#1}%
\providecommand \citenamefont [1]{#1}%
\providecommand \href@noop [0]{\@secondoftwo}%
\providecommand \href [0]{\begingroup \@sanitize@url \@href}%
\providecommand \@href[1]{\@@startlink{#1}\@@href}%
\providecommand \@@href[1]{\endgroup#1\@@endlink}%
\providecommand \@sanitize@url [0]{\catcode `\\12\catcode `\$12\catcode `\&12\catcode `\#12\catcode `\^12\catcode `\_12\catcode `\%12\relax}%
\providecommand \@@startlink[1]{}%
\providecommand \@@endlink[0]{}%
\providecommand \url  [0]{\begingroup\@sanitize@url \@url }%
\providecommand \@url [1]{\endgroup\@href {#1}{\urlprefix }}%
\providecommand \urlprefix  [0]{URL }%
\providecommand \Eprint [0]{\href }%
\providecommand \doibase [0]{https://doi.org/}%
\providecommand \selectlanguage [0]{\@gobble}%
\providecommand \bibinfo  [0]{\@secondoftwo}%
\providecommand \bibfield  [0]{\@secondoftwo}%
\providecommand \translation [1]{[#1]}%
\providecommand \BibitemOpen [0]{}%
\providecommand \bibitemStop [0]{}%
\providecommand \bibitemNoStop [0]{.\EOS\space}%
\providecommand \EOS [0]{\spacefactor3000\relax}%
\providecommand \BibitemShut  [1]{\csname bibitem#1\endcsname}%
\let\auto@bib@innerbib\@empty
\bibitem [{\citenamefont {Allen}\ \emph {et~al.}(1977)\citenamefont {Allen}, \citenamefont {Lucovsky},\ and\ \citenamefont {Mikkelsen}}]{AllenSSC77}%
  \BibitemOpen
  \bibfield  {author} {\bibinfo {author} {\bibfnamefont {J.}~\bibnamefont {Allen}}, \bibinfo {author} {\bibfnamefont {G.}~\bibnamefont {Lucovsky}},\ and\ \bibinfo {author} {\bibfnamefont {J.}~\bibnamefont {Mikkelsen}},\ }\bibfield  {title} {\bibinfo {title} {Optical properties and electronic structure of crossroads material {MnTe}},\ }\href {https://doi.org/https://doi.org/10.1016/0038-1098(77)90984-X} {\bibfield  {journal} {\bibinfo  {journal} {Solid State Commun.}\ }\textbf {\bibinfo {volume} {24}},\ \bibinfo {pages} {367} (\bibinfo {year} {1977})}\BibitemShut {NoStop}%
\bibitem [{\citenamefont {Mu}\ \emph {et~al.}(2019)\citenamefont {Mu}, \citenamefont {Hermann}, \citenamefont {Gorsse}, \citenamefont {Zhao}, \citenamefont {Manley}, \citenamefont {Fishman},\ and\ \citenamefont {Lindsay}}]{MuPRM19}%
  \BibitemOpen
  \bibfield  {author} {\bibinfo {author} {\bibfnamefont {S.}~\bibnamefont {Mu}}, \bibinfo {author} {\bibfnamefont {R.~P.}\ \bibnamefont {Hermann}}, \bibinfo {author} {\bibfnamefont {S.}~\bibnamefont {Gorsse}}, \bibinfo {author} {\bibfnamefont {H.}~\bibnamefont {Zhao}}, \bibinfo {author} {\bibfnamefont {M.~E.}\ \bibnamefont {Manley}}, \bibinfo {author} {\bibfnamefont {R.~S.}\ \bibnamefont {Fishman}},\ and\ \bibinfo {author} {\bibfnamefont {L.}~\bibnamefont {Lindsay}},\ }\bibfield  {title} {\bibinfo {title} {Phonons, magnons, and lattice thermal transport in antiferromagnetic semiconductor {MnTe}},\ }\href {https://doi.org/10.1103/PhysRevMaterials.3.025403} {\bibfield  {journal} {\bibinfo  {journal} {Phys. Rev. Mater.}\ }\textbf {\bibinfo {volume} {3}},\ \bibinfo {pages} {025403} (\bibinfo {year} {2019})}\BibitemShut {NoStop}%
\bibitem [{\citenamefont {Baltz}\ \emph {et~al.}(2018)\citenamefont {Baltz}, \citenamefont {Manchon}, \citenamefont {Tsoi}, \citenamefont {Moriyama}, \citenamefont {Ono},\ and\ \citenamefont {Tserkovnyak}}]{Baltz:2018_a}%
  \BibitemOpen
  \bibfield  {author} {\bibinfo {author} {\bibfnamefont {V.}~\bibnamefont {Baltz}}, \bibinfo {author} {\bibfnamefont {A.}~\bibnamefont {Manchon}}, \bibinfo {author} {\bibfnamefont {M.}~\bibnamefont {Tsoi}}, \bibinfo {author} {\bibfnamefont {T.}~\bibnamefont {Moriyama}}, \bibinfo {author} {\bibfnamefont {T.}~\bibnamefont {Ono}},\ and\ \bibinfo {author} {\bibfnamefont {Y.}~\bibnamefont {Tserkovnyak}},\ }\bibfield  {title} {\bibinfo {title} {Antiferromagnetic spintronics},\ }\href {https://doi.org/10.1103/RevModPhys.90.015005} {\bibfield  {journal} {\bibinfo  {journal} {Rev. Mod. Phys.}\ }\textbf {\bibinfo {volume} {90}},\ \bibinfo {pages} {015005} (\bibinfo {year} {2018})}\BibitemShut {NoStop}%
\bibitem [{\citenamefont {Kriegner}\ \emph {et~al.}(2016)\citenamefont {Kriegner}, \citenamefont {V{\'y}born{\'y}}, \citenamefont {Olejn{\'i}k}, \citenamefont {Reichlov{\'a}}, \citenamefont {Nov{\'a}k}, \citenamefont {Marti}, \citenamefont {Gazquez}, \citenamefont {Saidl}, \citenamefont {N{\v{e}}mec}, \citenamefont {Volobuev}, \citenamefont {Springholz}, \citenamefont {Hol{\'y}},\ and\ \citenamefont {Jungwirth}}]{KriegnerNC16}%
  \BibitemOpen
  \bibfield  {author} {\bibinfo {author} {\bibfnamefont {D.}~\bibnamefont {Kriegner}}, \bibinfo {author} {\bibfnamefont {K.}~\bibnamefont {V{\'y}born{\'y}}}, \bibinfo {author} {\bibfnamefont {K.}~\bibnamefont {Olejn{\'i}k}}, \bibinfo {author} {\bibfnamefont {H.}~\bibnamefont {Reichlov{\'a}}}, \bibinfo {author} {\bibfnamefont {V.}~\bibnamefont {Nov{\'a}k}}, \bibinfo {author} {\bibfnamefont {X.}~\bibnamefont {Marti}}, \bibinfo {author} {\bibfnamefont {J.}~\bibnamefont {Gazquez}}, \bibinfo {author} {\bibfnamefont {V.}~\bibnamefont {Saidl}}, \bibinfo {author} {\bibfnamefont {P.}~\bibnamefont {N{\v{e}}mec}}, \bibinfo {author} {\bibfnamefont {V.~V.}\ \bibnamefont {Volobuev}}, \bibinfo {author} {\bibfnamefont {G.}~\bibnamefont {Springholz}}, \bibinfo {author} {\bibfnamefont {V.}~\bibnamefont {Hol{\'y}}},\ and\ \bibinfo {author} {\bibfnamefont {T.}~\bibnamefont {Jungwirth}},\ }\bibfield  {title} {\bibinfo {title} {Multiple-stable anisotropic magnetoresistance memory in antiferromagnetic {MnTe}},\ }\href
  {https://doi.org/10.1038/ncomms11623} {\bibfield  {journal} {\bibinfo  {journal} {Nat. Commun.}\ }\textbf {\bibinfo {volume} {7}},\ \bibinfo {pages} {11623} (\bibinfo {year} {2016})}\BibitemShut {NoStop}%
\bibitem [{\citenamefont {Kriegner}\ \emph {et~al.}(2017)\citenamefont {Kriegner}, \citenamefont {Reichlova}, \citenamefont {Grenzer}, \citenamefont {Schmidt}, \citenamefont {Ressouche}, \citenamefont {Godinho}, \citenamefont {Wagner}, \citenamefont {Martin}, \citenamefont {Shick}, \citenamefont {Volobuev}, \citenamefont {Springholz}, \citenamefont {Hol\'y}, \citenamefont {Wunderlich}, \citenamefont {Jungwirth},\ and\ \citenamefont {V\'yborn\'y}}]{KriegnerPRB17}%
  \BibitemOpen
  \bibfield  {author} {\bibinfo {author} {\bibfnamefont {D.}~\bibnamefont {Kriegner}}, \bibinfo {author} {\bibfnamefont {H.}~\bibnamefont {Reichlova}}, \bibinfo {author} {\bibfnamefont {J.}~\bibnamefont {Grenzer}}, \bibinfo {author} {\bibfnamefont {W.}~\bibnamefont {Schmidt}}, \bibinfo {author} {\bibfnamefont {E.}~\bibnamefont {Ressouche}}, \bibinfo {author} {\bibfnamefont {J.}~\bibnamefont {Godinho}}, \bibinfo {author} {\bibfnamefont {T.}~\bibnamefont {Wagner}}, \bibinfo {author} {\bibfnamefont {S.~Y.}\ \bibnamefont {Martin}}, \bibinfo {author} {\bibfnamefont {A.~B.}\ \bibnamefont {Shick}}, \bibinfo {author} {\bibfnamefont {V.~V.}\ \bibnamefont {Volobuev}}, \bibinfo {author} {\bibfnamefont {G.}~\bibnamefont {Springholz}}, \bibinfo {author} {\bibfnamefont {V.}~\bibnamefont {Hol\'y}}, \bibinfo {author} {\bibfnamefont {J.}~\bibnamefont {Wunderlich}}, \bibinfo {author} {\bibfnamefont {T.}~\bibnamefont {Jungwirth}},\ and\ \bibinfo {author} {\bibfnamefont {K.}~\bibnamefont {V\'yborn\'y}},\ }\bibfield  {title}
  {\bibinfo {title} {Magnetic anisotropy in antiferromagnetic hexagonal {MnTe}},\ }\href {https://doi.org/10.1103/PhysRevB.96.214418} {\bibfield  {journal} {\bibinfo  {journal} {Phys. Rev. B}\ }\textbf {\bibinfo {volume} {96}},\ \bibinfo {pages} {214418} (\bibinfo {year} {2017})}\BibitemShut {NoStop}%
\bibitem [{\citenamefont {Kluczyk}\ \emph {et~al.}(2024)\citenamefont {Kluczyk}, \citenamefont {Gas}, \citenamefont {Grzybowski}, \citenamefont {Skupi\ifmmode~\acute{n}\else \'{n}\fi{}ski}, \citenamefont {Borysiewicz}, \citenamefont {Fas}, \citenamefont {Suffczy\ifmmode~\acute{n}\else \'{n}\fi{}ski}, \citenamefont {Domagala}, \citenamefont {Grasza}, \citenamefont {Mycielski}, \citenamefont {Baj}, \citenamefont {Ahn}, \citenamefont {V\'yborn\'y}, \citenamefont {Sawicki},\ and\ \citenamefont {Gryglas-Borysiewicz}}]{KluczykPRB24}%
  \BibitemOpen
  \bibfield  {author} {\bibinfo {author} {\bibfnamefont {K.~P.}\ \bibnamefont {Kluczyk}}, \bibinfo {author} {\bibfnamefont {K.}~\bibnamefont {Gas}}, \bibinfo {author} {\bibfnamefont {M.~J.}\ \bibnamefont {Grzybowski}}, \bibinfo {author} {\bibfnamefont {P.}~\bibnamefont {Skupi\ifmmode~\acute{n}\else \'{n}\fi{}ski}}, \bibinfo {author} {\bibfnamefont {M.~A.}\ \bibnamefont {Borysiewicz}}, \bibinfo {author} {\bibfnamefont {T.}~\bibnamefont {Fas}}, \bibinfo {author} {\bibfnamefont {J.}~\bibnamefont {Suffczy\ifmmode~\acute{n}\else \'{n}\fi{}ski}}, \bibinfo {author} {\bibfnamefont {J.~Z.}\ \bibnamefont {Domagala}}, \bibinfo {author} {\bibfnamefont {K.}~\bibnamefont {Grasza}}, \bibinfo {author} {\bibfnamefont {A.}~\bibnamefont {Mycielski}}, \bibinfo {author} {\bibfnamefont {M.}~\bibnamefont {Baj}}, \bibinfo {author} {\bibfnamefont {K.~H.}\ \bibnamefont {Ahn}}, \bibinfo {author} {\bibfnamefont {K.}~\bibnamefont {V\'yborn\'y}}, \bibinfo {author} {\bibfnamefont {M.}~\bibnamefont {Sawicki}},\ and\ \bibinfo {author}
  {\bibfnamefont {M.}~\bibnamefont {Gryglas-Borysiewicz}},\ }\bibfield  {title} {\bibinfo {title} {Coexistence of anomalous {Hall} effect and weak magnetization in a nominally collinear antiferromagnet {MnTe}},\ }\href {https://doi.org/10.1103/PhysRevB.110.155201} {\bibfield  {journal} {\bibinfo  {journal} {Phys. Rev. B}\ }\textbf {\bibinfo {volume} {110}},\ \bibinfo {pages} {155201} (\bibinfo {year} {2024})}\BibitemShut {NoStop}%
\bibitem [{\citenamefont {Hubert}\ \emph {et~al.}()\citenamefont {Hubert}, \citenamefont {Malecek}, \citenamefont {Ahn}, \citenamefont {Misek}, \citenamefont {Zelezny}, \citenamefont {Maca}, \citenamefont {Springholz}, \citenamefont {Veis},\ and\ \citenamefont {Vyborny}}]{Hubertpss25}%
  \BibitemOpen
  \bibfield  {author} {\bibinfo {author} {\bibfnamefont {M.}~\bibnamefont {Hubert}}, \bibinfo {author} {\bibfnamefont {T.}~\bibnamefont {Malecek}}, \bibinfo {author} {\bibfnamefont {K.-H.}\ \bibnamefont {Ahn}}, \bibinfo {author} {\bibfnamefont {M.}~\bibnamefont {Misek}}, \bibinfo {author} {\bibfnamefont {J.}~\bibnamefont {Zelezny}}, \bibinfo {author} {\bibfnamefont {F.}~\bibnamefont {Maca}}, \bibinfo {author} {\bibfnamefont {G.}~\bibnamefont {Springholz}}, \bibinfo {author} {\bibfnamefont {M.}~\bibnamefont {Veis}},\ and\ \bibinfo {author} {\bibfnamefont {K.}~\bibnamefont {Vyborny}},\ }\bibfield  {title} {\bibinfo {title} {Anomalous spectroscopical effects in an antiferromagnetic semiconductor: The case of magneto-optical kerr effect},\ }\href {https://doi.org/https://doi.org/10.1002/pssb.202400541} {\bibfield  {journal} {\bibinfo  {journal} {physica status solidi (b)}\ }\textbf {\bibinfo {volume} {n/a}},\ \bibinfo {pages} {2400541}}\BibitemShut {NoStop}%
\bibitem [{\citenamefont {Hariki}\ \emph {et~al.}(2024)\citenamefont {Hariki}, \citenamefont {Dal~Din}, \citenamefont {Amin}, \citenamefont {Yamaguchi}, \citenamefont {Badura}, \citenamefont {Kriegner}, \citenamefont {Edmonds}, \citenamefont {Campion}, \citenamefont {Wadley}, \citenamefont {Backes}, \citenamefont {Veiga}, \citenamefont {Dhesi}, \citenamefont {Springholz}, \citenamefont {\ifmmode~\check{S}\else \v{S}\fi{}mejkal}, \citenamefont {V\'yborn\'y}, \citenamefont {Jungwirth},\ and\ \citenamefont {Kune\ifmmode~\check{s}\else \v{s}\fi{}}}]{HarikiPRL24}%
  \BibitemOpen
  \bibfield  {author} {\bibinfo {author} {\bibfnamefont {A.}~\bibnamefont {Hariki}}, \bibinfo {author} {\bibfnamefont {A.}~\bibnamefont {Dal~Din}}, \bibinfo {author} {\bibfnamefont {O.~J.}\ \bibnamefont {Amin}}, \bibinfo {author} {\bibfnamefont {T.}~\bibnamefont {Yamaguchi}}, \bibinfo {author} {\bibfnamefont {A.}~\bibnamefont {Badura}}, \bibinfo {author} {\bibfnamefont {D.}~\bibnamefont {Kriegner}}, \bibinfo {author} {\bibfnamefont {K.~W.}\ \bibnamefont {Edmonds}}, \bibinfo {author} {\bibfnamefont {R.~P.}\ \bibnamefont {Campion}}, \bibinfo {author} {\bibfnamefont {P.}~\bibnamefont {Wadley}}, \bibinfo {author} {\bibfnamefont {D.}~\bibnamefont {Backes}}, \bibinfo {author} {\bibfnamefont {L.~S.~I.}\ \bibnamefont {Veiga}}, \bibinfo {author} {\bibfnamefont {S.~S.}\ \bibnamefont {Dhesi}}, \bibinfo {author} {\bibfnamefont {G.}~\bibnamefont {Springholz}}, \bibinfo {author} {\bibfnamefont {L.}~\bibnamefont {\ifmmode~\check{S}\else \v{S}\fi{}mejkal}}, \bibinfo {author} {\bibfnamefont {K.}~\bibnamefont {V\'yborn\'y}},
  \bibinfo {author} {\bibfnamefont {T.}~\bibnamefont {Jungwirth}},\ and\ \bibinfo {author} {\bibfnamefont {J.}~\bibnamefont {Kune\ifmmode~\check{s}\else \v{s}\fi{}}},\ }\bibfield  {title} {\bibinfo {title} {X-ray magnetic circular dichroism in altermagnetic {$\ensuremath{\alpha}$-MnTe}},\ }\href {https://doi.org/10.1103/PhysRevLett.132.176701} {\bibfield  {journal} {\bibinfo  {journal} {Phys. Rev. Lett.}\ }\textbf {\bibinfo {volume} {132}},\ \bibinfo {pages} {176701} (\bibinfo {year} {2024})}\BibitemShut {NoStop}%
\bibitem [{\citenamefont {Gonzalez~Betancourt}\ \emph {et~al.}(2024)\citenamefont {Gonzalez~Betancourt}, \citenamefont {Zub{\'a}{\v{c}}}, \citenamefont {Geishendorf}, \citenamefont {Ritzinger}, \citenamefont {R{\r{u}}{\v{z}}i{\v{c}}kov{\'a}}, \citenamefont {Kotte}, \citenamefont {{\v{Z}}elezn{\'y}}, \citenamefont {Olejn{\'i}k}, \citenamefont {Springholz}, \citenamefont {B{\"u}chner}, \citenamefont {Thomas}, \citenamefont {V{\'y}born{\'y}}, \citenamefont {Jungwirth}, \citenamefont {Reichlov{\'a}},\ and\ \citenamefont {Kriegner}}]{GonzalezBetancourtNJPS24}%
  \BibitemOpen
  \bibfield  {author} {\bibinfo {author} {\bibfnamefont {R.~D.}\ \bibnamefont {Gonzalez~Betancourt}}, \bibinfo {author} {\bibfnamefont {J.}~\bibnamefont {Zub{\'a}{\v{c}}}}, \bibinfo {author} {\bibfnamefont {K.}~\bibnamefont {Geishendorf}}, \bibinfo {author} {\bibfnamefont {P.}~\bibnamefont {Ritzinger}}, \bibinfo {author} {\bibfnamefont {B.}~\bibnamefont {R{\r{u}}{\v{z}}i{\v{c}}kov{\'a}}}, \bibinfo {author} {\bibfnamefont {T.}~\bibnamefont {Kotte}}, \bibinfo {author} {\bibfnamefont {J.}~\bibnamefont {{\v{Z}}elezn{\'y}}}, \bibinfo {author} {\bibfnamefont {K.}~\bibnamefont {Olejn{\'i}k}}, \bibinfo {author} {\bibfnamefont {G.}~\bibnamefont {Springholz}}, \bibinfo {author} {\bibfnamefont {B.}~\bibnamefont {B{\"u}chner}}, \bibinfo {author} {\bibfnamefont {A.}~\bibnamefont {Thomas}}, \bibinfo {author} {\bibfnamefont {K.}~\bibnamefont {V{\'y}born{\'y}}}, \bibinfo {author} {\bibfnamefont {T.}~\bibnamefont {Jungwirth}}, \bibinfo {author} {\bibfnamefont {H.}~\bibnamefont {Reichlov{\'a}}},\ and\ \bibinfo {author}
  {\bibfnamefont {D.}~\bibnamefont {Kriegner}},\ }\bibfield  {title} {\bibinfo {title} {Anisotropic magnetoresistance in altermagnetic {MnTe}},\ }\href {https://doi.org/10.1038/s44306-024-00046-z} {\bibfield  {journal} {\bibinfo  {journal} {npj Spintronics}\ }\textbf {\bibinfo {volume} {2}},\ \bibinfo {pages} {45} (\bibinfo {year} {2024})}\BibitemShut {NoStop}%
\bibitem [{\citenamefont {Nagaosa}\ \emph {et~al.}(2010)\citenamefont {Nagaosa}, \citenamefont {Sinova}, \citenamefont {Onoda}, \citenamefont {MacDonald},\ and\ \citenamefont {Ong}}]{Nagaosa:2010_a}%
  \BibitemOpen
  \bibfield  {author} {\bibinfo {author} {\bibfnamefont {N.}~\bibnamefont {Nagaosa}}, \bibinfo {author} {\bibfnamefont {J.}~\bibnamefont {Sinova}}, \bibinfo {author} {\bibfnamefont {S.}~\bibnamefont {Onoda}}, \bibinfo {author} {\bibfnamefont {A.~H.}\ \bibnamefont {MacDonald}},\ and\ \bibinfo {author} {\bibfnamefont {N.~P.}\ \bibnamefont {Ong}},\ }\bibfield  {title} {\bibinfo {title} {Anomalous {Hall} effect},\ }\href {https://doi.org/10.1103/RevModPhys.82.1539} {\bibfield  {journal} {\bibinfo  {journal} {Rev. Mod. Phys.}\ }\textbf {\bibinfo {volume} {82}},\ \bibinfo {pages} {1539} (\bibinfo {year} {2010})}\BibitemShut {NoStop}%
\bibitem [{\citenamefont {Ritzinger}\ and\ \citenamefont {Vyborny}(2023)}]{Ritzinger:2023_a}%
  \BibitemOpen
  \bibfield  {author} {\bibinfo {author} {\bibfnamefont {P.}~\bibnamefont {Ritzinger}}\ and\ \bibinfo {author} {\bibfnamefont {K.}~\bibnamefont {Vyborny}},\ }\bibfield  {title} {\bibinfo {title} {Anisotropic magnetoresistance: materials, models and applications},\ }\href {https://doi.org/10.1098/rsos.230564} {\bibfield  {journal} {\bibinfo  {journal} {R. Soc. Open Sci.}\ }\textbf {\bibinfo {volume} {10}},\ \bibinfo {pages} {230564} (\bibinfo {year} {2023})}\BibitemShut {NoStop}%
\bibitem [{\citenamefont {\ifmmode~\check{S}\else \v{S}\fi{}mejkal}\ \emph {et~al.}(2023)\citenamefont {\ifmmode~\check{S}\else \v{S}\fi{}mejkal}, \citenamefont {Marmodoro}, \citenamefont {Ahn}, \citenamefont {Gonz\'alez-Hern\'andez}, \citenamefont {Turek}, \citenamefont {Mankovsky}, \citenamefont {Ebert}, \citenamefont {D'Souza}, \citenamefont {\ifmmode~\check{S}\else \v{S}\fi{}ipr}, \citenamefont {Sinova},\ and\ \citenamefont {Jungwirth}}]{SmejkalPRL23}%
  \BibitemOpen
  \bibfield  {author} {\bibinfo {author} {\bibfnamefont {L.}~\bibnamefont {\ifmmode~\check{S}\else \v{S}\fi{}mejkal}}, \bibinfo {author} {\bibfnamefont {A.}~\bibnamefont {Marmodoro}}, \bibinfo {author} {\bibfnamefont {K.-H.}\ \bibnamefont {Ahn}}, \bibinfo {author} {\bibfnamefont {R.}~\bibnamefont {Gonz\'alez-Hern\'andez}}, \bibinfo {author} {\bibfnamefont {I.}~\bibnamefont {Turek}}, \bibinfo {author} {\bibfnamefont {S.}~\bibnamefont {Mankovsky}}, \bibinfo {author} {\bibfnamefont {H.}~\bibnamefont {Ebert}}, \bibinfo {author} {\bibfnamefont {S.~W.}\ \bibnamefont {D'Souza}}, \bibinfo {author} {\bibfnamefont {O.}~\bibnamefont {\ifmmode~\check{S}\else \v{S}\fi{}ipr}}, \bibinfo {author} {\bibfnamefont {J.}~\bibnamefont {Sinova}},\ and\ \bibinfo {author} {\bibfnamefont {T.}~\bibnamefont {Jungwirth}},\ }\bibfield  {title} {\bibinfo {title} {Chiral magnons in altermagnetic {${\mathrm{RuO}}_{2}$}},\ }\href {https://doi.org/10.1103/PhysRevLett.131.256703} {\bibfield  {journal} {\bibinfo  {journal} {Phys. Rev. Lett.}\
  }\textbf {\bibinfo {volume} {131}},\ \bibinfo {pages} {256703} (\bibinfo {year} {2023})}\BibitemShut {NoStop}%
\bibitem [{\citenamefont {Liu}\ \emph {et~al.}(2024)\citenamefont {Liu}, \citenamefont {Ozeki}, \citenamefont {Asai}, \citenamefont {Itoh},\ and\ \citenamefont {Masuda}}]{LiuPRL24}%
  \BibitemOpen
  \bibfield  {author} {\bibinfo {author} {\bibfnamefont {Z.}~\bibnamefont {Liu}}, \bibinfo {author} {\bibfnamefont {M.}~\bibnamefont {Ozeki}}, \bibinfo {author} {\bibfnamefont {S.}~\bibnamefont {Asai}}, \bibinfo {author} {\bibfnamefont {S.}~\bibnamefont {Itoh}},\ and\ \bibinfo {author} {\bibfnamefont {T.}~\bibnamefont {Masuda}},\ }\bibfield  {title} {\bibinfo {title} {Chiral split magnon in altermagnetic {MnTe}},\ }\href {https://doi.org/10.1103/PhysRevLett.133.156702} {\bibfield  {journal} {\bibinfo  {journal} {Phys. Rev. Lett.}\ }\textbf {\bibinfo {volume} {133}},\ \bibinfo {pages} {156702} (\bibinfo {year} {2024})}\BibitemShut {NoStop}%
\bibitem [{\citenamefont {\ifmmode~\check{S}\else \v{S}\fi{}mejkal}\ \emph {et~al.}(2022{\natexlab{a}})\citenamefont {\ifmmode~\check{S}\else \v{S}\fi{}mejkal}, \citenamefont {Sinova},\ and\ \citenamefont {Jungwirth}}]{SmejkalPRX22I}%
  \BibitemOpen
  \bibfield  {author} {\bibinfo {author} {\bibfnamefont {L.}~\bibnamefont {\ifmmode~\check{S}\else \v{S}\fi{}mejkal}}, \bibinfo {author} {\bibfnamefont {J.}~\bibnamefont {Sinova}},\ and\ \bibinfo {author} {\bibfnamefont {T.}~\bibnamefont {Jungwirth}},\ }\bibfield  {title} {\bibinfo {title} {Beyond conventional ferromagnetism and antiferromagnetism: A phase with nonrelativistic spin and crystal rotation symmetry},\ }\href {https://doi.org/10.1103/PhysRevX.12.031042} {\bibfield  {journal} {\bibinfo  {journal} {Phys. Rev. X}\ }\textbf {\bibinfo {volume} {12}},\ \bibinfo {pages} {031042} (\bibinfo {year} {2022}{\natexlab{a}})}\BibitemShut {NoStop}%
\bibitem [{\citenamefont {\ifmmode~\check{S}\else \v{S}\fi{}mejkal}\ \emph {et~al.}(2022{\natexlab{b}})\citenamefont {\ifmmode~\check{S}\else \v{S}\fi{}mejkal}, \citenamefont {Sinova},\ and\ \citenamefont {Jungwirth}}]{SmejkalPRX22II}%
  \BibitemOpen
  \bibfield  {author} {\bibinfo {author} {\bibfnamefont {L.}~\bibnamefont {\ifmmode~\check{S}\else \v{S}\fi{}mejkal}}, \bibinfo {author} {\bibfnamefont {J.}~\bibnamefont {Sinova}},\ and\ \bibinfo {author} {\bibfnamefont {T.}~\bibnamefont {Jungwirth}},\ }\bibfield  {title} {\bibinfo {title} {Emerging research landscape of altermagnetism},\ }\href {https://doi.org/10.1103/PhysRevX.12.040501} {\bibfield  {journal} {\bibinfo  {journal} {Phys. Rev. X}\ }\textbf {\bibinfo {volume} {12}},\ \bibinfo {pages} {040501} (\bibinfo {year} {2022}{\natexlab{b}})}\BibitemShut {NoStop}%
\bibitem [{\citenamefont {Liu}\ \emph {et~al.}(2025)\citenamefont {Liu}, \citenamefont {Dai},\ and\ \citenamefont {Bl{\"u}gel}}]{BluegelNP25}%
  \BibitemOpen
  \bibfield  {author} {\bibinfo {author} {\bibfnamefont {Q.}~\bibnamefont {Liu}}, \bibinfo {author} {\bibfnamefont {X.}~\bibnamefont {Dai}},\ and\ \bibinfo {author} {\bibfnamefont {S.}~\bibnamefont {Bl{\"u}gel}},\ }\bibfield  {title} {\bibinfo {title} {Different facets of unconventional magnetism},\ }\href {https://doi.org/10.1038/s41567-024-02750-3} {\bibfield  {journal} {\bibinfo  {journal} {Nature Phys.}\ }\textbf {\bibinfo {volume} {21}},\ \bibinfo {pages} {329} (\bibinfo {year} {2025})}\BibitemShut {NoStop}%
\bibitem [{Note1()}]{Note1}%
  \BibitemOpen
  \bibinfo {note} {Combined spatial and time inversion $T$. In addition to this symmetry, translations combined with $T$ also have to be absent among the symmetries. It should be stressed that non-collinear magnets can, of course, also have broken $PT$ symmetry.}\BibitemShut {Stop}%
\bibitem [{\citenamefont {Krempask{\'y}}\ \emph {et~al.}(2024)\citenamefont {Krempask{\'y}}, \citenamefont {{\v{S}}mejkal}, \citenamefont {D'Souza}, \citenamefont {Hajlaoui}, \citenamefont {Springholz}, \citenamefont {Uhl{\'i}{\v{r}}ov{\'a}}, \citenamefont {Alarab}, \citenamefont {Constantinou}, \citenamefont {Strocov}, \citenamefont {Usanov}, \citenamefont {Pudelko}, \citenamefont {Gonz{\'a}lez-Hern{\'a}ndez}, \citenamefont {Birk~Hellenes}, \citenamefont {Jansa}, \citenamefont {Reichlov{\'a}}, \citenamefont {{\v{S}}ob{\'a}{\v{n}}}, \citenamefont {Gonzalez~Betancourt}, \citenamefont {Wadley}, \citenamefont {Sinova}, \citenamefont {Kriegner}, \citenamefont {Min{\'a}r}, \citenamefont {Dil},\ and\ \citenamefont {Jungwirth}}]{KrempaskyNature24}%
  \BibitemOpen
  \bibfield  {author} {\bibinfo {author} {\bibfnamefont {J.}~\bibnamefont {Krempask{\'y}}}, \bibinfo {author} {\bibfnamefont {L.}~\bibnamefont {{\v{S}}mejkal}}, \bibinfo {author} {\bibfnamefont {S.~W.}\ \bibnamefont {D'Souza}}, \bibinfo {author} {\bibfnamefont {M.}~\bibnamefont {Hajlaoui}}, \bibinfo {author} {\bibfnamefont {G.}~\bibnamefont {Springholz}}, \bibinfo {author} {\bibfnamefont {K.}~\bibnamefont {Uhl{\'i}{\v{r}}ov{\'a}}}, \bibinfo {author} {\bibfnamefont {F.}~\bibnamefont {Alarab}}, \bibinfo {author} {\bibfnamefont {P.~C.}\ \bibnamefont {Constantinou}}, \bibinfo {author} {\bibfnamefont {V.}~\bibnamefont {Strocov}}, \bibinfo {author} {\bibfnamefont {D.}~\bibnamefont {Usanov}}, \bibinfo {author} {\bibfnamefont {W.~R.}\ \bibnamefont {Pudelko}}, \bibinfo {author} {\bibfnamefont {R.}~\bibnamefont {Gonz{\'a}lez-Hern{\'a}ndez}}, \bibinfo {author} {\bibfnamefont {A.}~\bibnamefont {Birk~Hellenes}}, \bibinfo {author} {\bibfnamefont {Z.}~\bibnamefont {Jansa}}, \bibinfo {author} {\bibfnamefont {H.}~\bibnamefont
  {Reichlov{\'a}}}, \bibinfo {author} {\bibfnamefont {Z.}~\bibnamefont {{\v{S}}ob{\'a}{\v{n}}}}, \bibinfo {author} {\bibfnamefont {R.~D.}\ \bibnamefont {Gonzalez~Betancourt}}, \bibinfo {author} {\bibfnamefont {P.}~\bibnamefont {Wadley}}, \bibinfo {author} {\bibfnamefont {J.}~\bibnamefont {Sinova}}, \bibinfo {author} {\bibfnamefont {D.}~\bibnamefont {Kriegner}}, \bibinfo {author} {\bibfnamefont {J.}~\bibnamefont {Min{\'a}r}}, \bibinfo {author} {\bibfnamefont {J.~H.}\ \bibnamefont {Dil}},\ and\ \bibinfo {author} {\bibfnamefont {T.}~\bibnamefont {Jungwirth}},\ }\bibfield  {title} {\bibinfo {title} {Altermagnetic lifting of {Kramers} spin degeneracy},\ }\href {https://doi.org/10.1038/s41586-023-06907-7} {\bibfield  {journal} {\bibinfo  {journal} {Nature}\ }\textbf {\bibinfo {volume} {626}},\ \bibinfo {pages} {517} (\bibinfo {year} {2024})}\BibitemShut {NoStop}%
\bibitem [{\citenamefont {Osumi}\ \emph {et~al.}(2024)\citenamefont {Osumi}, \citenamefont {Souma}, \citenamefont {Aoyama}, \citenamefont {Yamauchi}, \citenamefont {Honma}, \citenamefont {Nakayama}, \citenamefont {Takahashi}, \citenamefont {Ohgushi},\ and\ \citenamefont {Sato}}]{OsumiPRB24}%
  \BibitemOpen
  \bibfield  {author} {\bibinfo {author} {\bibfnamefont {T.}~\bibnamefont {Osumi}}, \bibinfo {author} {\bibfnamefont {S.}~\bibnamefont {Souma}}, \bibinfo {author} {\bibfnamefont {T.}~\bibnamefont {Aoyama}}, \bibinfo {author} {\bibfnamefont {K.}~\bibnamefont {Yamauchi}}, \bibinfo {author} {\bibfnamefont {A.}~\bibnamefont {Honma}}, \bibinfo {author} {\bibfnamefont {K.}~\bibnamefont {Nakayama}}, \bibinfo {author} {\bibfnamefont {T.}~\bibnamefont {Takahashi}}, \bibinfo {author} {\bibfnamefont {K.}~\bibnamefont {Ohgushi}},\ and\ \bibinfo {author} {\bibfnamefont {T.}~\bibnamefont {Sato}},\ }\bibfield  {title} {\bibinfo {title} {Observation of a giant band splitting in altermagnetic {MnTe}},\ }\href {https://doi.org/10.1103/PhysRevB.109.115102} {\bibfield  {journal} {\bibinfo  {journal} {Phys. Rev. B}\ }\textbf {\bibinfo {volume} {109}},\ \bibinfo {pages} {115102} (\bibinfo {year} {2024})}\BibitemShut {NoStop}%
\bibitem [{\citenamefont {Ferrer-Roca}\ \emph {et~al.}(2000)\citenamefont {Ferrer-Roca}, \citenamefont {Segura}, \citenamefont {Reig},\ and\ \citenamefont {Mu\~noz}}]{Ferrer-RocaPRB00}%
  \BibitemOpen
  \bibfield  {author} {\bibinfo {author} {\bibfnamefont {C.}~\bibnamefont {Ferrer-Roca}}, \bibinfo {author} {\bibfnamefont {A.}~\bibnamefont {Segura}}, \bibinfo {author} {\bibfnamefont {C.}~\bibnamefont {Reig}},\ and\ \bibinfo {author} {\bibfnamefont {V.}~\bibnamefont {Mu\~noz}},\ }\bibfield  {title} {\bibinfo {title} {Temperature and pressure dependence of the optical absorption in hexagonal {MnTe}},\ }\href {https://doi.org/10.1103/PhysRevB.61.13679} {\bibfield  {journal} {\bibinfo  {journal} {Phys. Rev. B}\ }\textbf {\bibinfo {volume} {61}},\ \bibinfo {pages} {13679} (\bibinfo {year} {2000})}\BibitemShut {NoStop}%
\bibitem [{\citenamefont {Faria~Junior}\ \emph {et~al.}(2023)\citenamefont {Faria~Junior}, \citenamefont {de~Mare}, \citenamefont {Zollner}, \citenamefont {Ahn}, \citenamefont {Erlingsson}, \citenamefont {van Schilfgaarde},\ and\ \citenamefont {V\'yborn\'y}}]{JuniorPRB23}%
  \BibitemOpen
  \bibfield  {author} {\bibinfo {author} {\bibfnamefont {P.~E.}\ \bibnamefont {Faria~Junior}}, \bibinfo {author} {\bibfnamefont {K.~A.}\ \bibnamefont {de~Mare}}, \bibinfo {author} {\bibfnamefont {K.}~\bibnamefont {Zollner}}, \bibinfo {author} {\bibfnamefont {K.-h.}\ \bibnamefont {Ahn}}, \bibinfo {author} {\bibfnamefont {S.~I.}\ \bibnamefont {Erlingsson}}, \bibinfo {author} {\bibfnamefont {M.}~\bibnamefont {van Schilfgaarde}},\ and\ \bibinfo {author} {\bibfnamefont {K.}~\bibnamefont {V\'yborn\'y}},\ }\bibfield  {title} {\bibinfo {title} {Sensitivity of the {MnTe} valence band to the orientation of magnetic moments},\ }\href {https://doi.org/10.1103/PhysRevB.107.L100417} {\bibfield  {journal} {\bibinfo  {journal} {Phys. Rev. B}\ }\textbf {\bibinfo {volume} {107}},\ \bibinfo {pages} {L100417} (\bibinfo {year} {2023})}\BibitemShut {NoStop}%
\bibitem [{\citenamefont {Madelung}\ \emph {et~al.}(2000)\citenamefont {Madelung}, \citenamefont {R{\"o}ssler},\ and\ \citenamefont {Schulz}}]{Madelung2000}%
  \BibitemOpen
  \bibfield  {author} {\bibinfo {author} {\bibfnamefont {O.}~\bibnamefont {Madelung}}, \bibinfo {author} {\bibfnamefont {U.}~\bibnamefont {R{\"o}ssler}},\ and\ \bibinfo {author} {\bibfnamefont {M.}~\bibnamefont {Schulz}},\ }\href@noop {} {\emph {\bibinfo {title} {Non-Tetrahedrally Bonded Binary Compounds II: Supplement to {Vol.} {III/17g} (Print Version) Revised and Updated Edition of {Vol. III/17g} (CD-ROM).}}}\ (\bibinfo  {publisher} {Springer},\ \bibinfo {year} {2000})\BibitemShut {NoStop}%
\bibitem [{\citenamefont {Komatsubara}\ \emph {et~al.}(1963)\citenamefont {Komatsubara}, \citenamefont {Murakami},\ and\ \citenamefont {Hirahara}}]{TakemiJPSJ63}%
  \BibitemOpen
  \bibfield  {author} {\bibinfo {author} {\bibfnamefont {T.}~\bibnamefont {Komatsubara}}, \bibinfo {author} {\bibfnamefont {M.}~\bibnamefont {Murakami}},\ and\ \bibinfo {author} {\bibfnamefont {E.}~\bibnamefont {Hirahara}},\ }\bibfield  {title} {\bibinfo {title} {Magnetic properties of manganese telluride single crystals},\ }\href {https://doi.org/10.1143/JPSJ.18.356} {\bibfield  {journal} {\bibinfo  {journal} {J. Phys. Soc. Jpn.}\ }\textbf {\bibinfo {volume} {18}},\ \bibinfo {pages} {356} (\bibinfo {year} {1963})}\BibitemShut {NoStop}%
\bibitem [{\citenamefont {Alaei}\ \emph {et~al.}(2025)\citenamefont {Alaei}, \citenamefont {Sobieszczyk}, \citenamefont {Ptok}, \citenamefont {Rezaei}, \citenamefont {Oganov},\ and\ \citenamefont {Qaiumzadeh}}]{AlaeiPRB25}%
  \BibitemOpen
  \bibfield  {author} {\bibinfo {author} {\bibfnamefont {M.}~\bibnamefont {Alaei}}, \bibinfo {author} {\bibfnamefont {P.}~\bibnamefont {Sobieszczyk}}, \bibinfo {author} {\bibfnamefont {A.}~\bibnamefont {Ptok}}, \bibinfo {author} {\bibfnamefont {N.}~\bibnamefont {Rezaei}}, \bibinfo {author} {\bibfnamefont {A.~R.}\ \bibnamefont {Oganov}},\ and\ \bibinfo {author} {\bibfnamefont {A.}~\bibnamefont {Qaiumzadeh}},\ }\bibfield  {title} {\bibinfo {title} {Origin of $a$-type antiferromagnetism and chiral split magnons in altermagnetic $\ensuremath{\alpha}$-mnte},\ }\href {https://doi.org/10.1103/PhysRevB.111.104416} {\bibfield  {journal} {\bibinfo  {journal} {Phys. Rev. B}\ }\textbf {\bibinfo {volume} {111}},\ \bibinfo {pages} {104416} (\bibinfo {year} {2025})}\BibitemShut {NoStop}%
\bibitem [{\citenamefont {Szuszkiewicz}\ \emph {et~al.}(2006)\citenamefont {Szuszkiewicz}, \citenamefont {Dynowska}, \citenamefont {Witkowska},\ and\ \citenamefont {Hennion}}]{SzuszkiewiczPRB06}%
  \BibitemOpen
  \bibfield  {author} {\bibinfo {author} {\bibfnamefont {W.}~\bibnamefont {Szuszkiewicz}}, \bibinfo {author} {\bibfnamefont {E.}~\bibnamefont {Dynowska}}, \bibinfo {author} {\bibfnamefont {B.}~\bibnamefont {Witkowska}},\ and\ \bibinfo {author} {\bibfnamefont {B.}~\bibnamefont {Hennion}},\ }\bibfield  {title} {\bibinfo {title} {Spin-wave measurements on hexagonal $\mathrm{MnTe}$ of $\mathrm{NiAs}$-type structure by inelastic neutron scattering},\ }\href {https://doi.org/10.1103/PhysRevB.73.104403} {\bibfield  {journal} {\bibinfo  {journal} {Phys. Rev. B}\ }\textbf {\bibinfo {volume} {73}},\ \bibinfo {pages} {104403} (\bibinfo {year} {2006})}\BibitemShut {NoStop}%
\bibitem [{\citenamefont {Moriya}(1960)}]{MoriyaPR60}%
  \BibitemOpen
  \bibfield  {author} {\bibinfo {author} {\bibfnamefont {T.}~\bibnamefont {Moriya}},\ }\bibfield  {title} {\bibinfo {title} {Theory of magnetism of {Ni${\mathrm{F}}_{2}$}},\ }\href {https://doi.org/10.1103/PhysRev.117.635} {\bibfield  {journal} {\bibinfo  {journal} {Phys. Rev.}\ }\textbf {\bibinfo {volume} {117}},\ \bibinfo {pages} {635} (\bibinfo {year} {1960})}\BibitemShut {NoStop}%
\bibitem [{\citenamefont {{O'Grady}}\ \emph {et~al.}(2020)\citenamefont {{O'Grady}}, \citenamefont {Sinclair}, \citenamefont {Elphick}, \citenamefont {Carpenter}, \citenamefont {Vallejo-Fernandez}, \citenamefont {Probert},\ and\ \citenamefont {Hirohata}}]{OGradyJAP20}%
  \BibitemOpen
  \bibfield  {author} {\bibinfo {author} {\bibfnamefont {K.}~\bibnamefont {{O'Grady}}}, \bibinfo {author} {\bibfnamefont {J.}~\bibnamefont {Sinclair}}, \bibinfo {author} {\bibfnamefont {K.}~\bibnamefont {Elphick}}, \bibinfo {author} {\bibfnamefont {R.}~\bibnamefont {Carpenter}}, \bibinfo {author} {\bibfnamefont {G.}~\bibnamefont {Vallejo-Fernandez}}, \bibinfo {author} {\bibfnamefont {M.~I.~J.}\ \bibnamefont {Probert}},\ and\ \bibinfo {author} {\bibfnamefont {A.}~\bibnamefont {Hirohata}},\ }\bibfield  {title} {\bibinfo {title} {Anisotropy in antiferromagnets},\ }\href {https://doi.org/10.1063/5.0006077} {\bibfield  {journal} {\bibinfo  {journal} {J. Appl. Phys.}\ }\textbf {\bibinfo {volume} {128}},\ \bibinfo {pages} {040901} (\bibinfo {year} {2020})}\BibitemShut {NoStop}%
\bibitem [{\citenamefont {Rezende}\ \emph {et~al.}(2019)\citenamefont {Rezende}, \citenamefont {Azevedo},\ and\ \citenamefont {Rodriguez-Suarez}}]{RezendeJAP19}%
  \BibitemOpen
  \bibfield  {author} {\bibinfo {author} {\bibfnamefont {S.~M.}\ \bibnamefont {Rezende}}, \bibinfo {author} {\bibfnamefont {A.}~\bibnamefont {Azevedo}},\ and\ \bibinfo {author} {\bibfnamefont {R.~L.}\ \bibnamefont {Rodriguez-Suarez}},\ }\bibfield  {title} {\bibinfo {title} {{Introduction to antiferromagnetic magnons}},\ }\href {https://doi.org/10.1063/1.5109132} {\bibfield  {journal} {\bibinfo  {journal} {J. Appl. Phys.}\ }\textbf {\bibinfo {volume} {126}},\ \bibinfo {pages} {151101} (\bibinfo {year} {2019})}\BibitemShut {NoStop}%
\bibitem [{\citenamefont {Kittel}(1951)}]{KittelPR51}%
  \BibitemOpen
  \bibfield  {author} {\bibinfo {author} {\bibfnamefont {C.}~\bibnamefont {Kittel}},\ }\bibfield  {title} {\bibinfo {title} {Theory of antiferromagnetic resonance},\ }\href {https://doi.org/10.1103/PhysRev.82.565} {\bibfield  {journal} {\bibinfo  {journal} {Phys. Rev.}\ }\textbf {\bibinfo {volume} {82}},\ \bibinfo {pages} {565} (\bibinfo {year} {1951})}\BibitemShut {NoStop}%
\bibitem [{\citenamefont {Keffer}\ and\ \citenamefont {Kittel}(1952)}]{KefferPR52}%
  \BibitemOpen
  \bibfield  {author} {\bibinfo {author} {\bibfnamefont {F.}~\bibnamefont {Keffer}}\ and\ \bibinfo {author} {\bibfnamefont {C.}~\bibnamefont {Kittel}},\ }\bibfield  {title} {\bibinfo {title} {Theory of antiferromagnetic resonance},\ }\href {https://doi.org/10.1103/PhysRev.85.329} {\bibfield  {journal} {\bibinfo  {journal} {Phys. Rev.}\ }\textbf {\bibinfo {volume} {85}},\ \bibinfo {pages} {329} (\bibinfo {year} {1952})}\BibitemShut {NoStop}%
\bibitem [{\citenamefont {Garcia-Gaitan}\ \emph {et~al.}(2025)\citenamefont {Garcia-Gaitan}, \citenamefont {Kefayati}, \citenamefont {Xiao},\ and\ \citenamefont {Nikoli\ifmmode~\acute{c}\else \'{c}\fi{}}}]{GarciaGaitan25}%
  \BibitemOpen
  \bibfield  {author} {\bibinfo {author} {\bibfnamefont {F.}~\bibnamefont {Garcia-Gaitan}}, \bibinfo {author} {\bibfnamefont {A.}~\bibnamefont {Kefayati}}, \bibinfo {author} {\bibfnamefont {J.~Q.}\ \bibnamefont {Xiao}},\ and\ \bibinfo {author} {\bibfnamefont {B.~K.}\ \bibnamefont {Nikoli\ifmmode~\acute{c}\else \'{c}\fi{}}},\ }\bibfield  {title} {\bibinfo {title} {Magnon spectrum of altermagnets beyond linear spin wave theory: Magnon-magnon interactions via time-dependent matrix product states versus atomistic spin dynamics},\ }\href {https://doi.org/10.1103/PhysRevB.111.L020407} {\bibfield  {journal} {\bibinfo  {journal} {Phys. Rev. B}\ }\textbf {\bibinfo {volume} {111}},\ \bibinfo {pages} {L020407} (\bibinfo {year} {2025})}\BibitemShut {NoStop}%
\bibitem [{\citenamefont {{Corr\^ea}}\ and\ \citenamefont {{V\'yborn\'y}}(2018)}]{Correa:2018_a}%
  \BibitemOpen
  \bibfield  {author} {\bibinfo {author} {\bibfnamefont {C.~A.}\ \bibnamefont {{Corr\^ea}}}\ and\ \bibinfo {author} {\bibfnamefont {K.}~\bibnamefont {{V\'yborn\'y}}},\ }\bibfield  {title} {\bibinfo {title} {Electronic structure and magnetic anisotropies of antiferromagnetic transition-metal difluorides},\ }\href {https://doi.org/10.1103/PhysRevB.97.235111} {\bibfield  {journal} {\bibinfo  {journal} {Phys. Rev. B}\ }\textbf {\bibinfo {volume} {97}},\ \bibinfo {pages} {235111} (\bibinfo {year} {2018})}\BibitemShut {NoStop}%
\bibitem [{\citenamefont {Bey}\ \emph {et~al.}(2024)\citenamefont {Bey}, \citenamefont {Fields}, \citenamefont {Combs}, \citenamefont {Markus}, \citenamefont {Beke}, \citenamefont {Wang}, \citenamefont {Ievlev}, \citenamefont {Zhukovskyi}, \citenamefont {Orlova}, \citenamefont {Forro}, \citenamefont {Bennett}, \citenamefont {Liu},\ and\ \citenamefont {Assaf}}]{BeyCM24}%
  \BibitemOpen
  \bibfield  {author} {\bibinfo {author} {\bibfnamefont {S.}~\bibnamefont {Bey}}, \bibinfo {author} {\bibfnamefont {S.~S.}\ \bibnamefont {Fields}}, \bibinfo {author} {\bibfnamefont {N.~G.}\ \bibnamefont {Combs}}, \bibinfo {author} {\bibfnamefont {B.~G.}\ \bibnamefont {Markus}}, \bibinfo {author} {\bibfnamefont {D.}~\bibnamefont {Beke}}, \bibinfo {author} {\bibfnamefont {J.}~\bibnamefont {Wang}}, \bibinfo {author} {\bibfnamefont {A.~V.}\ \bibnamefont {Ievlev}}, \bibinfo {author} {\bibfnamefont {M.}~\bibnamefont {Zhukovskyi}}, \bibinfo {author} {\bibfnamefont {T.}~\bibnamefont {Orlova}}, \bibinfo {author} {\bibfnamefont {L.}~\bibnamefont {Forro}}, \bibinfo {author} {\bibfnamefont {S.~P.}\ \bibnamefont {Bennett}}, \bibinfo {author} {\bibfnamefont {X.}~\bibnamefont {Liu}},\ and\ \bibinfo {author} {\bibfnamefont {B.~A.}\ \bibnamefont {Assaf}},\ }\href {https://arxiv.org/abs/2409.04567} {\bibinfo {title} {Unexpected tuning of the anomalous {Hall} effect in altermagnetic {MnTe} thin films}} (\bibinfo {year}
  {2024}),\ \Eprint {https://arxiv.org/abs/2409.04567} {arXiv:2409.04567 [cond-mat.mtrl-sci]} \BibitemShut {NoStop}%
\bibitem [{\citenamefont {T.~Nagamiya}\ and\ \citenamefont {Kubo}(1955)}]{NagamiyaAiP55}%
  \BibitemOpen
  \bibfield  {author} {\bibinfo {author} {\bibfnamefont {K.~Y.}\ \bibnamefont {T.~Nagamiya}}\ and\ \bibinfo {author} {\bibfnamefont {R.}~\bibnamefont {Kubo}},\ }\bibfield  {title} {\bibinfo {title} {Antiferromagnetism},\ }\href {https://doi.org/10.1080/00018735500101154} {\bibfield  {journal} {\bibinfo  {journal} {Adv. Phys.}\ }\textbf {\bibinfo {volume} {4}},\ \bibinfo {pages} {1} (\bibinfo {year} {1955})}\BibitemShut {NoStop}%
\bibitem [{\citenamefont {Jana}\ \emph {et~al.}(2023)\citenamefont {Jana}, \citenamefont {Kapuscinski}, \citenamefont {Mohelsky}, \citenamefont {Vaclavkova}, \citenamefont {Breslavetz}, \citenamefont {Orlita}, \citenamefont {Faugeras},\ and\ \citenamefont {Potemski}}]{JanaPRB23}%
  \BibitemOpen
  \bibfield  {author} {\bibinfo {author} {\bibfnamefont {D.}~\bibnamefont {Jana}}, \bibinfo {author} {\bibfnamefont {P.}~\bibnamefont {Kapuscinski}}, \bibinfo {author} {\bibfnamefont {I.}~\bibnamefont {Mohelsky}}, \bibinfo {author} {\bibfnamefont {D.}~\bibnamefont {Vaclavkova}}, \bibinfo {author} {\bibfnamefont {I.}~\bibnamefont {Breslavetz}}, \bibinfo {author} {\bibfnamefont {M.}~\bibnamefont {Orlita}}, \bibinfo {author} {\bibfnamefont {C.}~\bibnamefont {Faugeras}},\ and\ \bibinfo {author} {\bibfnamefont {M.}~\bibnamefont {Potemski}},\ }\bibfield  {title} {\bibinfo {title} {Magnon gap excitations and spin-entangled optical transition in the van der {Waals} antiferromagnet {${\text{NiPS}}_{3}$}},\ }\href {https://doi.org/10.1103/PhysRevB.108.115149} {\bibfield  {journal} {\bibinfo  {journal} {Phys. Rev. B}\ }\textbf {\bibinfo {volume} {108}},\ \bibinfo {pages} {115149} (\bibinfo {year} {2023})}\BibitemShut {NoStop}%
\bibitem [{\citenamefont {K{\"o}bler}\ and\ \citenamefont {Hoser}(2010)}]{Kobler10}%
  \BibitemOpen
  \bibfield  {author} {\bibinfo {author} {\bibfnamefont {U.}~\bibnamefont {K{\"o}bler}}\ and\ \bibinfo {author} {\bibfnamefont {A.}~\bibnamefont {Hoser}},\ }\href@noop {} {\emph {\bibinfo {title} {Renormalization group theory: impact on experimental magnetism}}},\ Vol.\ \bibinfo {volume} {127}\ (\bibinfo  {publisher} {Springer Science \& Business Media},\ \bibinfo {year} {2010})\BibitemShut {NoStop}%
\bibitem [{\citenamefont {Szuszkiewicz}\ \emph {et~al.}(1997)\citenamefont {Szuszkiewicz}, \citenamefont {Mohrange}, \citenamefont {Jouanne}, \citenamefont {Kanehisa}, \citenamefont {{\'S}wirkowicz}, \citenamefont {Dynowska}, \citenamefont {Janik}, \citenamefont {Wojtowicz},\ and\ \citenamefont {Kossut}}]{SzuszkiewiczAPP97}%
  \BibitemOpen
  \bibfield  {author} {\bibinfo {author} {\bibfnamefont {W.}~\bibnamefont {Szuszkiewicz}}, \bibinfo {author} {\bibfnamefont {J.}~\bibnamefont {Mohrange}}, \bibinfo {author} {\bibfnamefont {M.}~\bibnamefont {Jouanne}}, \bibinfo {author} {\bibfnamefont {M.}~\bibnamefont {Kanehisa}}, \bibinfo {author} {\bibfnamefont {R.}~\bibnamefont {{\'S}wirkowicz}}, \bibinfo {author} {\bibfnamefont {E.}~\bibnamefont {Dynowska}}, \bibinfo {author} {\bibfnamefont {E.}~\bibnamefont {Janik}}, \bibinfo {author} {\bibfnamefont {T.}~\bibnamefont {Wojtowicz}},\ and\ \bibinfo {author} {\bibfnamefont {J.}~\bibnamefont {Kossut}},\ }\bibfield  {title} {\bibinfo {title} {Temperature dependence of {Raman} scattering by magnons in bulk-like {MBE}-grown {MnTe} films},\ }\href {https://doi.org/https://doi.org/10.12693/APhysPolA.92.1025} {\bibfield  {journal} {\bibinfo  {journal} {Acta Phys. Pol. A}\ }\textbf {\bibinfo {volume} {92}},\ \bibinfo {pages} {1025} (\bibinfo {year} {1997})}\BibitemShut {NoStop}%
\bibitem [{\citenamefont {Szuszkiewicz}\ \emph {et~al.}(2014)\citenamefont {Szuszkiewicz}, \citenamefont {Jouanne}, \citenamefont {Morhange}, \citenamefont {Kanehisa}, \citenamefont {Dynowska}, \citenamefont {Gas}, \citenamefont {Janik}, \citenamefont {Karczewski}, \citenamefont {Kuna},\ and\ \citenamefont {Wojtowicz}}]{Szuszkiewiczpssb14}%
  \BibitemOpen
  \bibfield  {author} {\bibinfo {author} {\bibfnamefont {W.}~\bibnamefont {Szuszkiewicz}}, \bibinfo {author} {\bibfnamefont {M.}~\bibnamefont {Jouanne}}, \bibinfo {author} {\bibfnamefont {J.-F.}\ \bibnamefont {Morhange}}, \bibinfo {author} {\bibfnamefont {M.}~\bibnamefont {Kanehisa}}, \bibinfo {author} {\bibfnamefont {E.}~\bibnamefont {Dynowska}}, \bibinfo {author} {\bibfnamefont {K.}~\bibnamefont {Gas}}, \bibinfo {author} {\bibfnamefont {E.}~\bibnamefont {Janik}}, \bibinfo {author} {\bibfnamefont {G.}~\bibnamefont {Karczewski}}, \bibinfo {author} {\bibfnamefont {R.}~\bibnamefont {Kuna}},\ and\ \bibinfo {author} {\bibfnamefont {T.}~\bibnamefont {Wojtowicz}},\ }\bibfield  {title} {\bibinfo {title} {Raman scattering as a tool to characterize semiconductor crystals, thin layers, and low-dimensional structures containing transition metals},\ }\href {https://doi.org/https://doi.org/10.1002/pssb.201350142} {\bibfield  {journal} {\bibinfo  {journal} {phys. stat. sol. (b)}\ }\textbf {\bibinfo {volume} {251}},\
  \bibinfo {pages} {1133} (\bibinfo {year} {2014})}\BibitemShut {NoStop}%
\bibitem [{\citenamefont {Nagai}(1969)}]{NagaiPR69}%
  \BibitemOpen
  \bibfield  {author} {\bibinfo {author} {\bibfnamefont {O.}~\bibnamefont {Nagai}},\ }\bibfield  {title} {\bibinfo {title} {Theory of temperature-dependent magnon energies in antiferromagnets},\ }\href {https://doi.org/10.1103/PhysRev.180.557} {\bibfield  {journal} {\bibinfo  {journal} {Phys. Rev.}\ }\textbf {\bibinfo {volume} {180}},\ \bibinfo {pages} {557} (\bibinfo {year} {1969})}\BibitemShut {NoStop}%
\bibitem [{\citenamefont {Zhang}\ \emph {et~al.}(2021)\citenamefont {Zhang}, \citenamefont {Bialek}, \citenamefont {Magrez}, \citenamefont {Yu},\ and\ \citenamefont {Ansermet}}]{ZhangJMMM21}%
  \BibitemOpen
  \bibfield  {author} {\bibinfo {author} {\bibfnamefont {J.}~\bibnamefont {Zhang}}, \bibinfo {author} {\bibfnamefont {M.}~\bibnamefont {Bialek}}, \bibinfo {author} {\bibfnamefont {A.}~\bibnamefont {Magrez}}, \bibinfo {author} {\bibfnamefont {H.}~\bibnamefont {Yu}},\ and\ \bibinfo {author} {\bibfnamefont {J.-P.}\ \bibnamefont {Ansermet}},\ }\bibfield  {title} {\bibinfo {title} {Antiferromagnetic resonance in {TmFeO$_3$} at high temperatures},\ }\href {https://doi.org/https://doi.org/10.1016/j.jmmm.2020.167562} {\bibfield  {journal} {\bibinfo  {journal} {J. Magn. Magn. Mater.}\ }\textbf {\bibinfo {volume} {523}},\ \bibinfo {pages} {167562} (\bibinfo {year} {2021})}\BibitemShut {NoStop}%
\bibitem [{\citenamefont {{K\"obler}}\ \emph {et~al.}(2005)\citenamefont {{K\"obler}}, \citenamefont {Hoser},\ and\ \citenamefont {{Sch\"afer}}}]{KoblerPBCM05}%
  \BibitemOpen
  \bibfield  {author} {\bibinfo {author} {\bibfnamefont {U.}~\bibnamefont {{K\"obler}}}, \bibinfo {author} {\bibfnamefont {A.}~\bibnamefont {Hoser}},\ and\ \bibinfo {author} {\bibfnamefont {W.}~\bibnamefont {{Sch\"afer}}},\ }\bibfield  {title} {\bibinfo {title} {On the temperature dependence of the magnetic excitations},\ }\href {https://doi.org/https://doi.org/10.1016/j.physb.2005.03.035} {\bibfield  {journal} {\bibinfo  {journal} {Phys. B: Condens. Matter.}\ }\textbf {\bibinfo {volume} {364}},\ \bibinfo {pages} {55} (\bibinfo {year} {2005})}\BibitemShut {NoStop}%
\bibitem [{\citenamefont {Ashcroft}\ and\ \citenamefont {Mermin}(1976)}]{Ashcroft76}%
  \BibitemOpen
  \bibfield  {author} {\bibinfo {author} {\bibfnamefont {N.~W.}\ \bibnamefont {Ashcroft}}\ and\ \bibinfo {author} {\bibfnamefont {N.~D.}\ \bibnamefont {Mermin}},\ }\href@noop {} {\emph {\bibinfo {title} {{S}olid {S}tate {P}hysics}}}\ (\bibinfo  {publisher} {Holt-Saunders},\ \bibinfo {year} {1976})\BibitemShut {NoStop}%
\bibitem [{\citenamefont {Kotthaus}\ \emph {et~al.}(1975)\citenamefont {Kotthaus}, \citenamefont {Abstreiter}, \citenamefont {Koch},\ and\ \citenamefont {Ranvaud}}]{KotthausPRL75}%
  \BibitemOpen
  \bibfield  {author} {\bibinfo {author} {\bibfnamefont {J.~P.}\ \bibnamefont {Kotthaus}}, \bibinfo {author} {\bibfnamefont {G.}~\bibnamefont {Abstreiter}}, \bibinfo {author} {\bibfnamefont {J.~F.}\ \bibnamefont {Koch}},\ and\ \bibinfo {author} {\bibfnamefont {R.}~\bibnamefont {Ranvaud}},\ }\bibfield  {title} {\bibinfo {title} {Cyclotron resonance of localized electrons on a {Si} surface},\ }\href {https://doi.org/10.1103/PhysRevLett.34.151} {\bibfield  {journal} {\bibinfo  {journal} {Phys. Rev. Lett.}\ }\textbf {\bibinfo {volume} {34}},\ \bibinfo {pages} {151} (\bibinfo {year} {1975})}\BibitemShut {NoStop}%
\bibitem [{\citenamefont {Wiggins}\ \emph {et~al.}(1990)\citenamefont {Wiggins}, \citenamefont {Nicholas}, \citenamefont {Harris},\ and\ \citenamefont {Foxon}}]{WigginsSS90}%
  \BibitemOpen
  \bibfield  {author} {\bibinfo {author} {\bibfnamefont {G.}~\bibnamefont {Wiggins}}, \bibinfo {author} {\bibfnamefont {R.}~\bibnamefont {Nicholas}}, \bibinfo {author} {\bibfnamefont {J.}~\bibnamefont {Harris}},\ and\ \bibinfo {author} {\bibfnamefont {C.}~\bibnamefont {Foxon}},\ }\bibfield  {title} {\bibinfo {title} {Bound state cyclotron resonance in modulation doped {GaAs/AlGaAs} quantum wells},\ }\href {https://doi.org/https://doi.org/10.1016/0039-6028(90)90937-4} {\bibfield  {journal} {\bibinfo  {journal} {Surf. Sci.}\ }\textbf {\bibinfo {volume} {229}},\ \bibinfo {pages} {488} (\bibinfo {year} {1990})}\BibitemShut {NoStop}%
\bibitem [{\citenamefont {Seck}\ \emph {et~al.}(1995)\citenamefont {Seck}, \citenamefont {Potemski}, \citenamefont {Huant}, \citenamefont {Wyder},\ and\ \citenamefont {Weimann}}]{SeckPBCM95}%
  \BibitemOpen
  \bibfield  {author} {\bibinfo {author} {\bibfnamefont {M.}~\bibnamefont {Seck}}, \bibinfo {author} {\bibfnamefont {M.}~\bibnamefont {Potemski}}, \bibinfo {author} {\bibfnamefont {S.}~\bibnamefont {Huant}}, \bibinfo {author} {\bibfnamefont {P.}~\bibnamefont {Wyder}},\ and\ \bibinfo {author} {\bibfnamefont {G.}~\bibnamefont {Weimann}},\ }\bibfield  {title} {\bibinfo {title} {Cyclotron resonance of low concentration {2D} electron gases in {GaAs/AlGaAs} heterostructures},\ }\href {https://doi.org/https://doi.org/10.1016/0921-4526(94)01096-J} {\bibfield  {journal} {\bibinfo  {journal} {Phys. B: Condens. Matter.}\ }\textbf {\bibinfo {volume} {211}},\ \bibinfo {pages} {470} (\bibinfo {year} {1995})}\BibitemShut {NoStop}%
\bibitem [{\citenamefont {Oshikiri}\ \emph {et~al.}(2001)\citenamefont {Oshikiri}, \citenamefont {Imanaka}, \citenamefont {Aryasetiawan},\ and\ \citenamefont {Kido}}]{OshikiriPBCM01}%
  \BibitemOpen
  \bibfield  {author} {\bibinfo {author} {\bibfnamefont {M.}~\bibnamefont {Oshikiri}}, \bibinfo {author} {\bibfnamefont {Y.}~\bibnamefont {Imanaka}}, \bibinfo {author} {\bibfnamefont {F.}~\bibnamefont {Aryasetiawan}},\ and\ \bibinfo {author} {\bibfnamefont {G.}~\bibnamefont {Kido}},\ }\bibfield  {title} {\bibinfo {title} {Comparison of the electron effective mass of the $n$-type {ZnO} in the wurtzite structure measured by cyclotron resonance and calculated from first principle theory},\ }\href {https://doi.org/https://doi.org/10.1016/S0921-4526(01)00365-9} {\bibfield  {journal} {\bibinfo  {journal} {Phys. B: Condens. Matter.}\ }\textbf {\bibinfo {volume} {298}},\ \bibinfo {pages} {472} (\bibinfo {year} {2001})},\ \bibinfo {note} {international Conference on High Magnetic Fields in Semiconductors}\BibitemShut {NoStop}%
\bibitem [{\citenamefont {Zanmarchi}(1967)}]{ZanmarchiJPCS67}%
  \BibitemOpen
  \bibfield  {author} {\bibinfo {author} {\bibfnamefont {G.}~\bibnamefont {Zanmarchi}},\ }\bibfield  {title} {\bibinfo {title} {Optical measurements on the antiferromagnetic semiconductor {MnTe}},\ }\href {https://doi.org/https://doi.org/10.1016/0022-3697(67)90235-1} {\bibfield  {journal} {\bibinfo  {journal} {J. Phys. Chem. Solids}\ }\textbf {\bibinfo {volume} {28}},\ \bibinfo {pages} {2123} (\bibinfo {year} {1967})}\BibitemShut {NoStop}%
\bibitem [{\citenamefont {Zanmarchi}\ and\ \citenamefont {Haas}(1968)}]{ZanmarchiJAP68}%
  \BibitemOpen
  \bibfield  {author} {\bibinfo {author} {\bibfnamefont {G.}~\bibnamefont {Zanmarchi}}\ and\ \bibinfo {author} {\bibfnamefont {C.}~\bibnamefont {Haas}},\ }\bibfield  {title} {\bibinfo {title} {Magnon drag at optical frequencies and the infrared spectrum of {MnTe}},\ }\href {https://doi.org/10.1063/1.2163537} {\bibfield  {journal} {\bibinfo  {journal} {J. Appl. Phys.}\ }\textbf {\bibinfo {volume} {39}},\ \bibinfo {pages} {596} (\bibinfo {year} {1968})}\BibitemShut {NoStop}%
\bibitem [{\citenamefont {Yumnam}\ \emph {et~al.}(2024)\citenamefont {Yumnam}, \citenamefont {Moseley}, \citenamefont {Paddison}, \citenamefont {Suggs}, \citenamefont {Zappala}, \citenamefont {Parker}, \citenamefont {Granroth}, \citenamefont {Morris}, \citenamefont {Polash}, \citenamefont {Vashaee}, \citenamefont {McGuire}, \citenamefont {Zhao}, \citenamefont {Manley}, \citenamefont {Frandsen},\ and\ \citenamefont {Hermann}}]{YumnamPRB24}%
  \BibitemOpen
  \bibfield  {author} {\bibinfo {author} {\bibfnamefont {G.}~\bibnamefont {Yumnam}}, \bibinfo {author} {\bibfnamefont {D.~H.}\ \bibnamefont {Moseley}}, \bibinfo {author} {\bibfnamefont {J.~A.~M.}\ \bibnamefont {Paddison}}, \bibinfo {author} {\bibfnamefont {C.~Z.}\ \bibnamefont {Suggs}}, \bibinfo {author} {\bibfnamefont {E.}~\bibnamefont {Zappala}}, \bibinfo {author} {\bibfnamefont {D.~S.}\ \bibnamefont {Parker}}, \bibinfo {author} {\bibfnamefont {G.~E.}\ \bibnamefont {Granroth}}, \bibinfo {author} {\bibfnamefont {G.~D.}\ \bibnamefont {Morris}}, \bibinfo {author} {\bibfnamefont {M.~M.~H.}\ \bibnamefont {Polash}}, \bibinfo {author} {\bibfnamefont {D.}~\bibnamefont {Vashaee}}, \bibinfo {author} {\bibfnamefont {M.~A.}\ \bibnamefont {McGuire}}, \bibinfo {author} {\bibfnamefont {H.}~\bibnamefont {Zhao}}, \bibinfo {author} {\bibfnamefont {M.~E.}\ \bibnamefont {Manley}}, \bibinfo {author} {\bibfnamefont {B.~A.}\ \bibnamefont {Frandsen}},\ and\ \bibinfo {author} {\bibfnamefont {R.~P.}\ \bibnamefont {Hermann}},\
  }\bibfield  {title} {\bibinfo {title} {Magnon gap tuning in lithium-doped {MnTe}},\ }\href {https://doi.org/10.1103/PhysRevB.109.214434} {\bibfield  {journal} {\bibinfo  {journal} {Phys. Rev. B}\ }\textbf {\bibinfo {volume} {109}},\ \bibinfo {pages} {214434} (\bibinfo {year} {2024})}\BibitemShut {NoStop}%
\bibitem [{\citenamefont {Moseley}\ \emph {et~al.}(2022)\citenamefont {Moseley}, \citenamefont {Taddei}, \citenamefont {Yan}, \citenamefont {McGuire}, \citenamefont {Calder}, \citenamefont {Polash}, \citenamefont {Vashaee}, \citenamefont {Zhang}, \citenamefont {Zhao}, \citenamefont {Parker}, \citenamefont {Fishman},\ and\ \citenamefont {Hermann}}]{MoseleyPRB22}%
  \BibitemOpen
  \bibfield  {author} {\bibinfo {author} {\bibfnamefont {D.~H.}\ \bibnamefont {Moseley}}, \bibinfo {author} {\bibfnamefont {K.~M.}\ \bibnamefont {Taddei}}, \bibinfo {author} {\bibfnamefont {J.}~\bibnamefont {Yan}}, \bibinfo {author} {\bibfnamefont {M.~A.}\ \bibnamefont {McGuire}}, \bibinfo {author} {\bibfnamefont {S.}~\bibnamefont {Calder}}, \bibinfo {author} {\bibfnamefont {M.~M.~H.}\ \bibnamefont {Polash}}, \bibinfo {author} {\bibfnamefont {D.}~\bibnamefont {Vashaee}}, \bibinfo {author} {\bibfnamefont {X.}~\bibnamefont {Zhang}}, \bibinfo {author} {\bibfnamefont {H.}~\bibnamefont {Zhao}}, \bibinfo {author} {\bibfnamefont {D.~S.}\ \bibnamefont {Parker}}, \bibinfo {author} {\bibfnamefont {R.~S.}\ \bibnamefont {Fishman}},\ and\ \bibinfo {author} {\bibfnamefont {R.~P.}\ \bibnamefont {Hermann}},\ }\bibfield  {title} {\bibinfo {title} {Giant doping response of magnetic anisotropy in mnte},\ }\href {https://doi.org/10.1103/PhysRevMaterials.6.014404} {\bibfield  {journal} {\bibinfo  {journal} {Phys. Rev. Mater.}\
  }\textbf {\bibinfo {volume} {6}},\ \bibinfo {pages} {014404} (\bibinfo {year} {2022})}\BibitemShut {NoStop}%
\bibitem [{\citenamefont {McCreary}\ \emph {et~al.}(2020)\citenamefont {McCreary}, \citenamefont {Simpson}, \citenamefont {Mai}, \citenamefont {McMichael}, \citenamefont {Douglas}, \citenamefont {Butch}, \citenamefont {Dennis}, \citenamefont {Vald\'es~Aguilar},\ and\ \citenamefont {Hight~Walker}}]{McCreary:2020_a}%
  \BibitemOpen
  \bibfield  {author} {\bibinfo {author} {\bibfnamefont {A.}~\bibnamefont {McCreary}}, \bibinfo {author} {\bibfnamefont {J.~R.}\ \bibnamefont {Simpson}}, \bibinfo {author} {\bibfnamefont {T.~T.}\ \bibnamefont {Mai}}, \bibinfo {author} {\bibfnamefont {R.~D.}\ \bibnamefont {McMichael}}, \bibinfo {author} {\bibfnamefont {J.~E.}\ \bibnamefont {Douglas}}, \bibinfo {author} {\bibfnamefont {N.}~\bibnamefont {Butch}}, \bibinfo {author} {\bibfnamefont {C.}~\bibnamefont {Dennis}}, \bibinfo {author} {\bibfnamefont {R.}~\bibnamefont {Vald\'es~Aguilar}},\ and\ \bibinfo {author} {\bibfnamefont {A.~R.}\ \bibnamefont {Hight~Walker}},\ }\bibfield  {title} {\bibinfo {title} {Quasi-two-dimensional magnon identification in antiferromagnetic {FePS}$_{3}$ via {magneto-Raman} spectroscopy},\ }\href {https://doi.org/10.1103/PhysRevB.101.064416} {\bibfield  {journal} {\bibinfo  {journal} {Phys. Rev. B}\ }\textbf {\bibinfo {volume} {101}},\ \bibinfo {pages} {064416} (\bibinfo {year} {2020})}\BibitemShut {NoStop}%
\bibitem [{\citenamefont {Csontosov\'a}\ \emph {et~al.}(2023)\citenamefont {Csontosov\'a}, \citenamefont {Chaloupka}, \citenamefont {Shinaoka}, \citenamefont {Hariki},\ and\ \citenamefont {Kune\ifmmode~\check{s}\else \v{s}\fi{}}}]{Csontosova:2023_a}%
  \BibitemOpen
  \bibfield  {author} {\bibinfo {author} {\bibfnamefont {D.}~\bibnamefont {Csontosov\'a}}, \bibinfo {author} {\bibfnamefont {J.}~\bibnamefont {Chaloupka}}, \bibinfo {author} {\bibfnamefont {H.}~\bibnamefont {Shinaoka}}, \bibinfo {author} {\bibfnamefont {A.}~\bibnamefont {Hariki}},\ and\ \bibinfo {author} {\bibfnamefont {J.}~\bibnamefont {Kune\ifmmode~\check{s}\else \v{s}\fi{}}},\ }\bibfield  {title} {\bibinfo {title} {Hidden covalent insulator and spin excitations in {SrRu}$_{2}${O}$_{6}$},\ }\href {https://doi.org/10.1103/PhysRevB.108.195137} {\bibfield  {journal} {\bibinfo  {journal} {Phys. Rev. B}\ }\textbf {\bibinfo {volume} {108}},\ \bibinfo {pages} {195137} (\bibinfo {year} {2023})}\BibitemShut {NoStop}%
\bibitem [{\citenamefont {Bey}\ \emph {et~al.}(2025)\citenamefont {Bey}, \citenamefont {Zhukovskyi}, \citenamefont {Orlova}, \citenamefont {Fields}, \citenamefont {Lauter}, \citenamefont {Ambaye}, \citenamefont {Ievlev}, \citenamefont {Bennett}, \citenamefont {Liu},\ and\ \citenamefont {Assaf}}]{BeyCM25}%
  \BibitemOpen
  \bibfield  {author} {\bibinfo {author} {\bibfnamefont {S.}~\bibnamefont {Bey}}, \bibinfo {author} {\bibfnamefont {M.}~\bibnamefont {Zhukovskyi}}, \bibinfo {author} {\bibfnamefont {T.}~\bibnamefont {Orlova}}, \bibinfo {author} {\bibfnamefont {S.}~\bibnamefont {Fields}}, \bibinfo {author} {\bibfnamefont {V.}~\bibnamefont {Lauter}}, \bibinfo {author} {\bibfnamefont {H.}~\bibnamefont {Ambaye}}, \bibinfo {author} {\bibfnamefont {A.}~\bibnamefont {Ievlev}}, \bibinfo {author} {\bibfnamefont {S.~P.}\ \bibnamefont {Bennett}}, \bibinfo {author} {\bibfnamefont {X.}~\bibnamefont {Liu}},\ and\ \bibinfo {author} {\bibfnamefont {B.~A.}\ \bibnamefont {Assaf}},\ }\href {https://arxiv.org/abs/2504.12126} {\bibinfo {title} {Interface, bulk and surface structure of heteroepitaxial altermagnetic {{$\alpha$}-MnTe} films grown on {GaAs(111)}}} (\bibinfo {year} {2025}),\ \Eprint {https://arxiv.org/abs/2504.12126} {arXiv:2504.12126 [cond-mat.mtrl-sci]} \BibitemShut {NoStop}%
\bibitem [{\citenamefont {Bialek}\ \emph {et~al.}(2022)\citenamefont {Bialek}, \citenamefont {Zhang}, \citenamefont {Yu},\ and\ \citenamefont {Ansermet}}]{BialekAPL22}%
  \BibitemOpen
  \bibfield  {author} {\bibinfo {author} {\bibfnamefont {M.}~\bibnamefont {Bialek}}, \bibinfo {author} {\bibfnamefont {J.}~\bibnamefont {Zhang}}, \bibinfo {author} {\bibfnamefont {H.}~\bibnamefont {Yu}},\ and\ \bibinfo {author} {\bibfnamefont {J.-P.}\ \bibnamefont {Ansermet}},\ }\bibfield  {title} {\bibinfo {title} {Antiferromagnetic resonance in {$\alpha$-Fe$_2$O$_3$} up to its {N\'eel} temperature},\ }\href {https://doi.org/10.1063/5.0094868} {\bibfield  {journal} {\bibinfo  {journal} {Appl. Phys. Lett.}\ }\textbf {\bibinfo {volume} {121}},\ \bibinfo {pages} {032401} (\bibinfo {year} {2022})}\BibitemShut {NoStop}%
\bibitem [{\citenamefont {Blumenschein}\ \emph {et~al.}(2020)\citenamefont {Blumenschein}, \citenamefont {Kadlec}, \citenamefont {Romanyuk}, \citenamefont {Paskova}, \citenamefont {Muth},\ and\ \citenamefont {Kadlec}}]{BlumenscheinJAP20}%
  \BibitemOpen
  \bibfield  {author} {\bibinfo {author} {\bibfnamefont {N.}~\bibnamefont {Blumenschein}}, \bibinfo {author} {\bibfnamefont {C.}~\bibnamefont {Kadlec}}, \bibinfo {author} {\bibfnamefont {O.}~\bibnamefont {Romanyuk}}, \bibinfo {author} {\bibfnamefont {T.}~\bibnamefont {Paskova}}, \bibinfo {author} {\bibfnamefont {J.~F.}\ \bibnamefont {Muth}},\ and\ \bibinfo {author} {\bibfnamefont {F.}~\bibnamefont {Kadlec}},\ }\bibfield  {title} {\bibinfo {title} {Dielectric and conducting properties of unintentionally and {Sn-doped} {$\beta$-Ga$_2$O$_3$} studied by terahertz spectroscopy},\ }\href {https://doi.org/10.1063/1.5143735} {\bibfield  {journal} {\bibinfo  {journal} {J. Appl. Phys.}\ }\textbf {\bibinfo {volume} {127}},\ \bibinfo {pages} {165702} (\bibinfo {year} {2020})}\BibitemShut {NoStop}%
\bibitem [{\citenamefont {{Tesa\v r}}\ \emph {et~al.}(2018)\citenamefont {{Tesa\v r}}, \citenamefont {{\v Sindler}}, \citenamefont {{Kol\'a\v cek}},\ and\ \citenamefont {Skrbek}}]{TesarRSI18}%
  \BibitemOpen
  \bibfield  {author} {\bibinfo {author} {\bibfnamefont {R.}~\bibnamefont {{Tesa\v r}}}, \bibinfo {author} {\bibfnamefont {M.}~\bibnamefont {{\v Sindler}}}, \bibinfo {author} {\bibfnamefont {J.}~\bibnamefont {{Kol\'a\v cek}}},\ and\ \bibinfo {author} {\bibfnamefont {L.}~\bibnamefont {Skrbek}},\ }\bibfield  {title} {\bibinfo {title} {{Terahertz wire-grid circular polarizer tuned by lock-in detection method}},\ }\href {https://doi.org/10.1063/1.5025427} {\bibfield  {journal} {\bibinfo  {journal} {Rev. Sci. Instrum.}\ }\textbf {\bibinfo {volume} {89}},\ \bibinfo {pages} {083114} (\bibinfo {year} {2018})}\BibitemShut {NoStop}%
\bibitem [{\citenamefont {Barra}\ \emph {et~al.}(2006)\citenamefont {Barra}, \citenamefont {Hassan}, \citenamefont {Janoschka}, \citenamefont {Schmidt},\ and\ \citenamefont {Sch{\"u}nemann}}]{BarraAMR06}%
  \BibitemOpen
  \bibfield  {author} {\bibinfo {author} {\bibfnamefont {A.~L.}\ \bibnamefont {Barra}}, \bibinfo {author} {\bibfnamefont {A.~K.}\ \bibnamefont {Hassan}}, \bibinfo {author} {\bibfnamefont {A.}~\bibnamefont {Janoschka}}, \bibinfo {author} {\bibfnamefont {C.~L.}\ \bibnamefont {Schmidt}},\ and\ \bibinfo {author} {\bibfnamefont {V.}~\bibnamefont {Sch{\"u}nemann}},\ }\bibfield  {title} {\bibinfo {title} {Broad-band quasi-optical {HF-EPR} spectroscopy: Application to the study of the ferrous iron center from a rubredoxin mutant},\ }\href {https://doi.org/10.1007/BF03166208} {\bibfield  {journal} {\bibinfo  {journal} {Appl. Magn. Reson.}\ }\textbf {\bibinfo {volume} {30}},\ \bibinfo {pages} {385} (\bibinfo {year} {2006})}\BibitemShut {NoStop}%
\bibitem [{\citenamefont {Mackay}\ and\ \citenamefont {Lakhtakia}(2022)}]{Mackay20}%
  \BibitemOpen
  \bibfield  {author} {\bibinfo {author} {\bibfnamefont {T.~G.}\ \bibnamefont {Mackay}}\ and\ \bibinfo {author} {\bibfnamefont {A.}~\bibnamefont {Lakhtakia}},\ }\href@noop {} {\emph {\bibinfo {title} {The transfer-matrix method in electromagnetics and optics}}}\ (\bibinfo  {publisher} {Springer Nature},\ \bibinfo {year} {2022})\BibitemShut {NoStop}%
\bibitem [{\citenamefont {Bia\l{}ek}\ \emph {et~al.}(2021)\citenamefont {Bia\l{}ek}, \citenamefont {Zhang}, \citenamefont {Yu},\ and\ \citenamefont {Ansermet}}]{Bialek21}%
  \BibitemOpen
  \bibfield  {author} {\bibinfo {author} {\bibfnamefont {M.}~\bibnamefont {Bia\l{}ek}}, \bibinfo {author} {\bibfnamefont {J.}~\bibnamefont {Zhang}}, \bibinfo {author} {\bibfnamefont {H.}~\bibnamefont {Yu}},\ and\ \bibinfo {author} {\bibfnamefont {J.-P.}\ \bibnamefont {Ansermet}},\ }\bibfield  {title} {\bibinfo {title} {Strong coupling of antiferromagnetic resonance with subterahertz cavity fields},\ }\href {https://doi.org/10.1103/PhysRevApplied.15.044018} {\bibfield  {journal} {\bibinfo  {journal} {Phys. Rev. Appl.}\ }\textbf {\bibinfo {volume} {15}},\ \bibinfo {pages} {044018} (\bibinfo {year} {2021})}\BibitemShut {NoStop}%
\bibitem [{\citenamefont {Boventer}\ \emph {et~al.}(2023)\citenamefont {Boventer}, \citenamefont {Simensen}, \citenamefont {Brekke}, \citenamefont {Weides}, \citenamefont {Anane}, \citenamefont {Kl\"aui}, \citenamefont {Brataas},\ and\ \citenamefont {Lebrun}}]{Boventer23}%
  \BibitemOpen
  \bibfield  {author} {\bibinfo {author} {\bibfnamefont {I.}~\bibnamefont {Boventer}}, \bibinfo {author} {\bibfnamefont {H.~T.}\ \bibnamefont {Simensen}}, \bibinfo {author} {\bibfnamefont {B.}~\bibnamefont {Brekke}}, \bibinfo {author} {\bibfnamefont {M.}~\bibnamefont {Weides}}, \bibinfo {author} {\bibfnamefont {A.}~\bibnamefont {Anane}}, \bibinfo {author} {\bibfnamefont {M.}~\bibnamefont {Kl\"aui}}, \bibinfo {author} {\bibfnamefont {A.}~\bibnamefont {Brataas}},\ and\ \bibinfo {author} {\bibfnamefont {R.}~\bibnamefont {Lebrun}},\ }\bibfield  {title} {\bibinfo {title} {Antiferromagnetic cavity magnon polaritons in collinear and canted phases of hematite},\ }\href {https://doi.org/10.1103/PhysRevApplied.19.014071} {\bibfield  {journal} {\bibinfo  {journal} {Phys. Rev. Appl.}\ }\textbf {\bibinfo {volume} {19}},\ \bibinfo {pages} {014071} (\bibinfo {year} {2023})}\BibitemShut {NoStop}%
\end{thebibliography}

\end{document}